\documentclass[]{IEEEtran}

\usepackage{graphicx}
\usepackage{color}
\usepackage{amssymb}
\usepackage{amsmath}
\usepackage{amsthm}
\usepackage{ifthen}

\newboolean{figurelist}
\setboolean{figurelist}{false}

\newcommand	
	{\incfig}	
	[3]	
	{%
	\ifthenelse{\boolean{figurelist}}
		{
		\immediate\write\outstream{cp fig-#1.pdf arXiv}
		}
		{}
	\begin{figure}[!t]
		\centering
			\includegraphics
				[width=#2\linewidth]	
				{fig-#1}
			\caption{#3}
			\label{fig:#1}	
	\end{figure}
	}

\newcommand	
	{\comment}	
	[1]
	{
	\begin{samepage}
	\begin{center}
	\framebox{
		\parbox{3in}{
		\textbf{Comment:}
			#1
		}
		}
	\end{center}
	\end{samepage}
	}
	
\newcommand 
	{\doctable}
	[5] 
	{
	\begin{table}[!t]
	
	\caption{#2}
	\label{tbl:#1}
	\centering
	\begin{tabular}{#3}
	#4
	\end{tabular}
	\end{table}
	}

\newenvironment{example}
	{
	\begin{LaTeXdescription}
	\item[Example:]
	}
	{
	\end{LaTeXdescription}
	}

\newcommand 
	{\citeref}
	[1] 
	{\cite{RefWorks:#1}}

\newcommand 
	{\citerefWloc}
	[2] 
	{\cite[#2]{RefWorks:#1}}
	
\newcommand 
	{\citereftwo}
	[2] 
	{\cite{RefWorks:#1, RefWorks:#2}}

\newcommand 
	{\citerefthree}
	[3] 
	{\cite{RefWorks:#1, RefWorks:#2, RefWorks:#3}}
	
\newcommand 
	{\citereffour}
	[4] 
	{\cite{RefWorks:#1, RefWorks:#2, RefWorks:#3, RefWorks:#4}}

\newcommand 
	{\citereffive}
	[5] 
	{\cite{
	RefWorks:#1, RefWorks:#2, RefWorks:#3, RefWorks:#4,RefWorks:#5}
	}
	
\newcommand 
	{\citerefsix}
	[6] 
	{\cite{
	RefWorks:#1, RefWorks:#2, RefWorks:#3,RefWorks:#4,RefWorks:#5,RefWorks:#6}
	}

\newenvironment{makefigurelist}
	{
	\ifthenelse{\boolean{figurelist}}
		{
		\newwrite\outstream
		\immediate\openout\outstream=figure_list
		\immediate\write\outstream{!/bin/bash}
		\immediate\write\outstream{}
		}
		{}
	}
	{
	\ifthenelse{\boolean{figurelist}}
		{
		\immediate\closeout\outstream
		}
		{}
	}
	
\newenvironment
	{changea}{\color{black}}{}%
\newenvironment
	{changeb}{\color{black}}{}%
\newenvironment
	{changec}{\color{black}}{}%
		
\newcommand{\convolve}{\otimes}						
\newcommand{\trans}[1]{{#1}^{\,\ensuremath{\mathsf{T}}}}           
\newcommand{\cc}[1]{{#1}^{\ast}}                            			
\newcommand{\hc}[1]{{#1}^{\,\dagger}}                            		

\newcommand{\energy}{\mathcal E}						

\newcommand{\power}{\mathcal P}						

\newcommand{\expec}[1]{\mathbb E \left[ #1 \right]}

\newcommand{\pe}{P_{\text{\,E}}}

\newcommand{\rate}{\mathcal R}

\newcommand{\npsd}{\mathcal N}
\newcommand{\ecode}{\energy_{\text{cod}}}
\newcommand{\ebit}{\energy_{\text{bit}}}
\newcommand{\ncode}{N_{\text{cod}}}
\newcommand{\bw}{\mathcal B}
\newcommand{\bcode}{B_{\text{cod}}}
\newcommand{\tcode}{T_{\text{cod}}}
\newcommand{\bwexpansion}{\gamma_{\text{exp}}}

\newcommand{\bcoherence}{B_{\text{coh}}}
\newcommand{\tcoherence}{T_{\text{coh}}}
\newcommand{\diff}{\mathrm d} 

\begin{document}

\title{Design for minimum energy\\
in starship and interstellar communication}%

\author{
David G. Messerschmitt
\thanks{Department of Electrical Engineering and Computer Sciences,
University of California at Berkeley.}%
\thanks{This work is licensed under a Creative Commons Attribution-NonCommercial-ShareAlike 3.0 Unported License,
http://creativecommons.org/licenses/by-nc-sa/3.0/.}%
\thanks{Citation:
	D.G. Messerschmitt, ''Design for minimum energy in starship and interstellar communication'', 
	submitted to the \emph{Journal of the British Interplanetary Society}
	(available at arxiv.org/abs/1402.1215).}%
}%

\maketitle

\begin{makefigurelist}

\begin{abstract}
\begin{changeb}
Microwave digital communication at interstellar distances applies to starship and extraterrestrial civilization (SETI and METI) communication. Large distances demand large transmitted power and/or large antennas, while the propagation is transparent over a wide bandwidth. Recognizing a fundamental tradeoff, reduced energy delivered to the receiver at the expense of wide bandwidth (the opposite of terrestrial objectives) is advantageous. Wide bandwidth also results in simpler design and implementation, allowing circumvention of dispersion and scattering arising in the interstellar medium and motion effects and obviating any related processing. The minimum energy delivered to the receiver per bit of information is determined by cosmic microwave background alone. By mapping a single bit onto a carrier burst, the Morse code invented for the telegraph in 1836 comes closer to this minimum energy than approaches used in modern terrestrial radio. Rather than the terrestrial approach of adding phases and amplitudes to increases information capacity while minimizing bandwidth, adding multiple time-frequency locations for carrier bursts increases capacity while minimizing energy per information bit. The resulting location code is extremely simple and yet can approach the minimum energy as bandwidth is expanded. It is consistent with easy discovery, since carrier bursts are energetic and straightforward modifications to post-detection pattern recognition can identify burst patterns. The interstellar coherence hole captures the time and frequency coherence constraints, and observations of the environment by transmitter and receiver constrain the burst parameters and limit the search scope.
\end{changeb}
\end{abstract}

\section{Introduction}

There has been considerable experience in communication with ''deep space'' probes 
in and around our solar system at radio wavelengths \citeref{660},
and there is growing interest in optical \citeref{661} as well.
Communication of information at much greater distances, such as with starships and
extraterrestrial civilizations, introduces new challenges, and has not been addressed
either theoretically or empirically.
Some of those challenges are 
addressed here from the perspective of communication engineering,
emphasizing radio (rather than optical) wavelengths.
\begin{changea}
The insights here are highly relevant to both
the search for extraterrestrial intelligence (SETI) by informing the discovery of information-bearing signals,
and messaging for extraterrestrial intelligence (METI), which transmits information-bearing signals.
\end{changea}
\begin{changec}
They are also relevant to the design of two-way links with interstellar spacecraft (often called starships).
\end{changec}

\doctable
	{acronyms}
	{Acronyms}
	{l p{6cm}}
	{
	\bfseries Acronym & \bfseries Definition 
	\\
	\hline
	AWGN & Additive white Gaussian noise
	\\
	CMB & Cosmic background radiation
	\\
	CSP & Coding and signal processing
	\\
	ICH & Interstellar coherence hole
	\\
	ISM & Interstellar medium
	\\
	SETI & Search for interstellar intelligence
	\\
	\hline
	}

\begin{changea}
This context is illustrated in Fig. \ref{fig:blockDiagram}, showing the subsystems of an
end-to-end system communicating information by radio.
This paper concerns the coding and signal processing (CSP) subsystem.
It is presumed that a message input is represented digitally
(composed of discrete symbols), 
in which case it can always be represented as a stream of bits.
The transmit-end CSP inputs information bits and outputs a baseband waveform
\begin{changec}
(the spectrum of which is concentrated about d.c.).
\end{changec}
The radio subsystem
\begin{changec}
(including modulation to passband, centered about a carrier frequency $f_c$,
\end{changec}
a radio-frequency transmitter, transmit/receive antennas, and demodulation
\begin{changec}
back to baseband)
\end{changec}
delivers a replica of this baseband waveform to the receive-end CSP
with impairments (such as distortion and noise) introduced by
the physical environment (the interstellar propagation and motion).
The modeling of these impairments, which profoundly affect the CSP,
is summarized here and addressed in greater depth elsewhere \citeref{664}.

\incfig
	{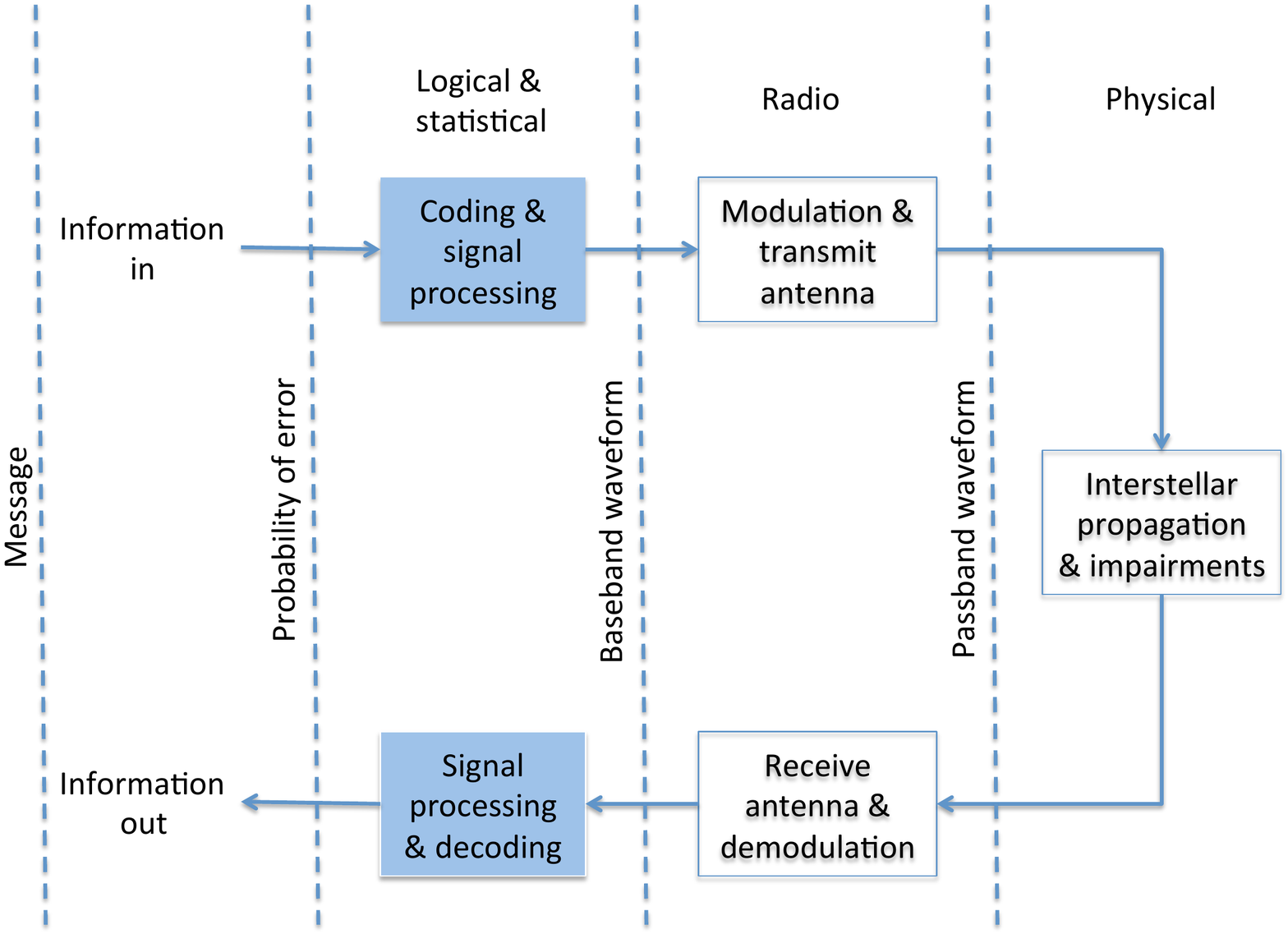}
	{1}
	{The major functional blocks in an interstellar communication system, with
	transmit on the top row and receive on the bottom row.
	These blocks are separated into (from left to right) information, baseband processing,
	passband radio, and physical propagation.
	}

The CSP realizes the essential function of mapping information bits into
a continuous-time baseband waveform suitable for transmission
as an electromagnetic wave.
The radio subsystem attempts to minimize impairments,
for example by reducing noise introduced in the receiver.
Through the choice of antennas, in the context of a given transmission distance
it determines the radiated power necessary to deliver
a needed level of power at receive baseband.
The remaining performance characteristics of the system
are determined by the CSP, including the fidelity of the message replica delivered by the
receive-end CSP
and the resources consumed in achieving this fidelity.
The fidelity is usually measured by probability of error, and
the primary resources of interest are the bandwidth of the radiated signal
and the signal power that must be delivered to receive baseband.
These resources can be manipulated over many orders of magnitude
through the design of the CSP, and thus it is the subsystem with the greatest opportunity
to manipulate the message fidelity and resources.

A complete messaging system combines the expertise of several disciplines,
including astronomy and astrophysics to model the physical, 
radio astronomy and electronics to design the radio,
and applied statistics to design the CSP.
The CSP has been the target of extensive research and practice in the
context of commercial terrestrial wireless systems, and that experience is
directly applicable to the interstellar case.
The goal here is to communicate to a wider audience
the opportunities afforded by the CSP.
Since the knowledge and techniques applied in the CSP may be unfamiliar to many readers,
the emphasis here is on reviewing results and developing intuition,
with details documented elsewhere \citeref{656}.

One of the tradeoffs determined by the CSP
is between bandwidth and received power; inevitably,
increasing either allows the other to be reduced.
\end{changea}
We argue that in communication at the large interstellar distances,
large propagation losses place a premium on minimizing energy requirements.
Other challenges applying specifically to communication with other civilizations are dispersive effects
arising in the interstellar medium (ISM) at distances greater than hundreds of light years,
relative motion, a lack of prior design coordination between
the two ends, and the challenges of discovering a signal with unknown design and parameterization.
It turns out that these considerations are closely linked, with both energy minimization
and interstellar impairments contributing beneficially to both
an implicit form of coordination between transmitter and receiver, as well as 
informing a discovery search.
\begin{changea}
In the case of starships, relativistic effects also become critical 
due to large relative velocities \citeref{665}.
\end{changea}

\section{Requirements}

There are two distinct applications of interstellar communication: 
communication with starships and with extraterrestrial civilizations. 
These two applications invoke distinctly different requirements significantly 
influencing the design.

\subsubsection{Starships}

Communication with a starship will be two-way, and the two ends of a communication link can be designed as a unit. 
Typical uses will be to send uplink control instructions to a starship, 
and to return performance parameters and scientific data on a downlink. 
Effectiveness in the control function is enhanced if the round-trip latency is minimized, 
but large speed-of-light propagation delays place a lower bound on this latency.
The only adjustable parameter is the uplink and downlink transmit times for a message, 
which is reduced with higher information rates. 
High downlink rates allow more scientific observations to be collected and returned to Earth. 
The accuracy of control and the integrity of scientific data demands reliability, in the form of a low error rate
and/or the ability to repeat messages on request.

Starships may travel at near the speed of light in order to reach the destination in reasonable time.
Thus relativistic effects are much more prominent than in communication with civilizations, which
are expected to be moving a much lower relative speeds.
\begin{changec}
Many propagation effects will be moderated or absent at the shorter distances expected to a starship.
\end{changec}

\subsubsection{Civilizations} 

Initial discovery of a signal from a civilization and establishment 
of two-way communication lacks coordination, and this
presents a difficult challenge \citeref{657}. 
Discovery of the signal absent any prior knowledge of its structure or parameterization is crucial. 

In the phase before two-way communication,
we/they are likely to carefully compose a message revealing something about our/their culture and state of knowledge. 
Composition of such a message should be a careful deliberative process, and changes to that message will probably occur infrequently, on timeframes of years or decades. 
Because we don`t know when and where such a message will be received, we/they are forced to transmit the message repeatedly. 
In this event, reliable reception (low error rate) for each instance of the message need not be a requirement, because a reliable rendition can be recovered from multiple unreliable replicas.
\begin{changec}
\begin{example}
For an average of one error in a thousand bits
($\pe = 10^{-3}$ where $\pe$ is the probability of error)
 for a single reception, after observing and combining five (or seven) replicas of a message on average only one out of 100 megabits (or 28 gigabits) will still be in error.
\end{example}
\end{changec}

Message transmission time is also not critical. Even in two-way communication.
the total round trip latency includes two message transmission times and two
propagation times, but the former will usually be insignificant relative to the latter.
For example, at a rate of one bit per second, 40 megabytes of message per decade is transmitted, but a decade is not particularly significant in the context of a propagation delay of  centuries or millennia.

At hundreds or thousands of light years, there are additional impairments
arising in the ISM to overcome at radio wavelengths, in the form of 
dispersion and scattering due to clouds of partially ionized gases. 
Pulsar astronomers and astrophysicists have observed and “reverse engineered” these impairments, 
providing a solid basis for design absent even the possibility of direct experimentation.

\subsubsection{Energy}

The distance from Earth to the nearest star is about 8900 times the distance to the outer planet Neptune, and a civilization at 1000 light years away is about $2 \times 10^6$ times as far away as Neptune.
\begin{changea}
For fixed antennas,
power or energy loss due to propagation is proportional to distance-squared,
resulting in a $7.8 \times 10^7$ and $4.4 \times 10^{12}$ increase in loss respectively.
\end{changea}
Propagation loss can be compensated by a larger antenna at transmitter and/or receiver, 
or by an increase in transmit power.
\begin{changea}
However this tradeoff is managed, overall costs are reduced by lowering the requirement
on power delivered at receiver baseband, as is the focus of our CSP design.
\end{changea}

To contact other civilizations,
an omnidirectional transmit antenna can address many targets at once \citeref{602},
but a transmitter seeking to conserve energy will sequentially scan targets 
using a highly directive antenna, especially in light of the signal characteristics described later.
\begin{changec}
Starships are likely to be much closer than even the nearest civilizations,
but the cost of either a large transmit antenna or transmit energy is likely to be considerably greater
on a spacecraft than on a planet.
\end{changec}
\begin{changea}
In both cases,
reducing the energy delivered to the receiver baseband is beneficial, and substantial reductions
are possible thru CSP design without compromising message fidelity.
\end{changea}

Another civilization may be much more advanced, and thus may employ means of communication that are inconceivable to us, 
or which can be hypothesized but are beyond our
technological capability or resources.
For example, a civilization may be able to exploit an exotic physics
that artificially reduces propagation distance, and such ''shortcuts'' through space-time
 reduce round-trip latency as well as propagation loss.
However, when Earth-based designers set out to build either an interstellar transmitter or
receiver, the design is constrained by our available technology.
This is not as constraining as it may appear, since in the context of quantum physics
as we know it and conventional radio propagation, today's technology on Earth can approach the fundamental
limit on delivered energy to the receiver.
Within this context
this limit cannot be circumvented by another civilization, no matter how advanced.

\section{Compatibility without coordination}

Even though one civilization is designing a transmitter and the other a receiver, 
the only hope of compatibility is for each to design an end-to-end system. 
That way, each contemplates and accounts for the challenges of the other. 
Even then there remains a lot of design freedom and a galaxy full of clever ideas, with many possibilities.
Such design freedom is particularly difficult to overcome in initial discovery of
a signal.
This challenge is addressed here by a four-prong approach illustrated in Fig. \ref{fig:approachingFL}:
\begin{itemize}
\item
Keep things as simple as possible without compromise in important objectives.
\item
Choose design objectives that are sensible in the context of interstellar communication.
In this paper the minimization of energy delivered to the receiver
is the overriding objective, given that this energy is directly or indirectly
(through antenna size) a major cost.
\item
Base the design on overcoming impairments due to the ISM
and relative transmitter-receiver motion that are observable by both
transmitter and receiver designers.
\item
Base the design on fundamental principles likely to be known to both transmitter
and receiver designers.
Specifically, this paper proposes minimizing the received energy
subject to the objective of reliable extraction of information from the signal.
\end{itemize}

\incfig
	{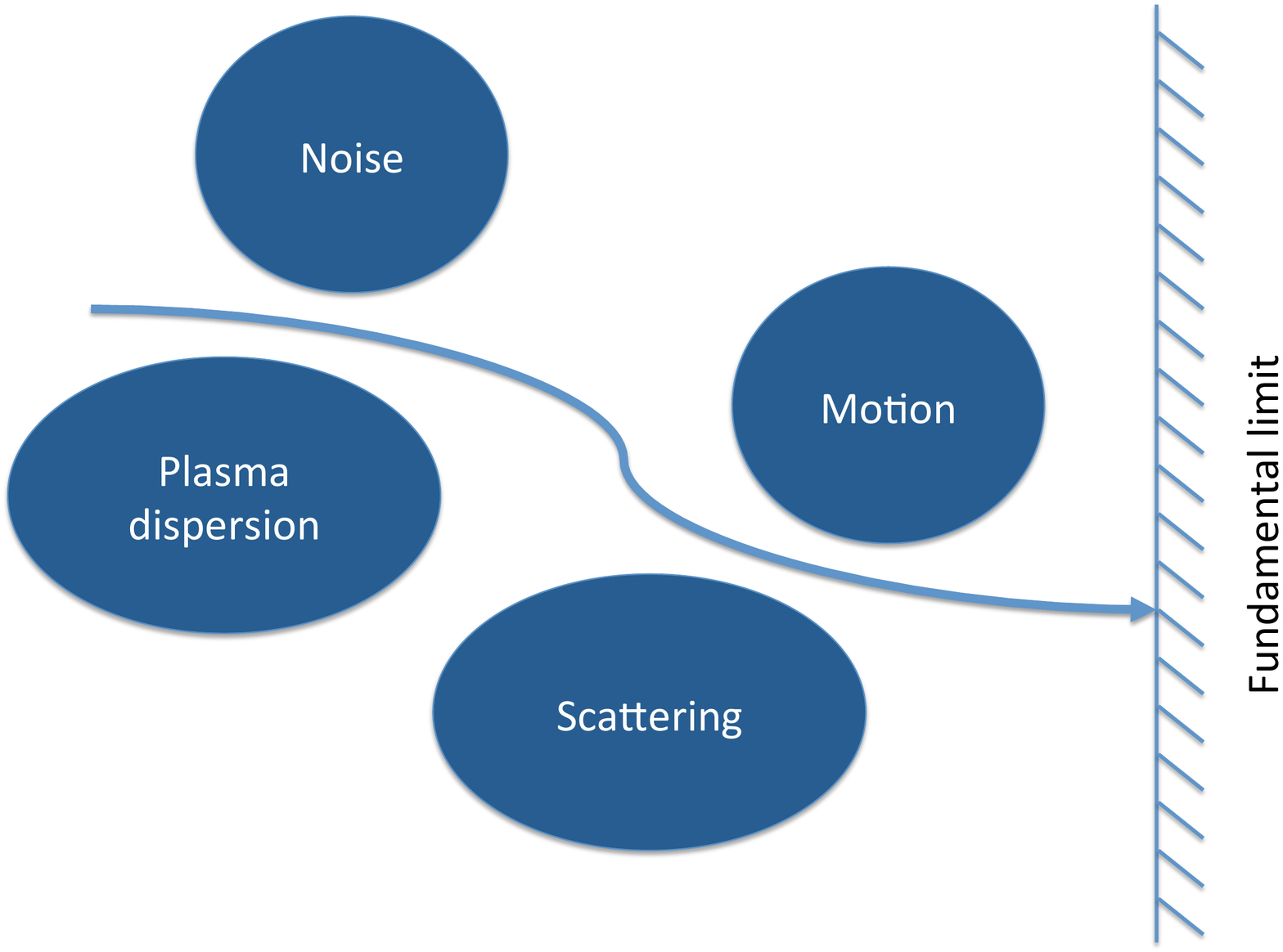}
	{.8}
	{A method for achieving implicit coordination by approaching a fundamental
	limit in the context of a set of mutually observable ISM and motion impairments.}

The simplicity argument postulates that complexity is an obstacle to finding common ground in the absence of coordination. Similar to Occam`s razor in philosophy, it seeks the simplest design that meets the needs and requirements of interstellar communication. Stated in a negative way, designers should avoid any gratuitous requirements that increase the complexity of the solution and fail to produce substantive advantage to either transmitter or receiver.

It is presumed that other civilizations have observed the ISM, and arrived at similar models of impairments to radio propagation originating there.
This is expected of any designer seeking to communicate successfully through the ISM.
For a civilization such as ours lacking galactic-scale probes,
direct experimentation is not possible but detailed information about the medium is available
indirectly through the pulsar observations by astronomers and modeling by
astrophysicists.

Communications is blessed with mathematically provable fundamental limits 
due originally to Shannon in 1948 \citeref{164}. 
Those limits, as well as means of approaching them, depend on both the nature of impairments introduced in the physical environment and the underlying performance objective. 
Thus, both minimum delivered energy and interstellar impairments are relevant.  
Since 1948 communications has been dominated by an unceasing effort to approach those fundamental limits, with good success based on advancing technology and conceptual advances.
The same would be expected of another civilization at least as technologically advanced as ours. If both the transmitter and receiver designers seek to approach fundamental limits, they will arrive at similar design principles and mutually benefit from the resulting performance advantages.

As will be shown,
all three elements (simplicity, minimum energy, and interstellar impairments) work
constructively and collectively to drastically narrow the signal characteristics.

\section{Fundamental limit: Gaussian noise}
\label{sec:energyLimitGaussian}

Cosmic microwave background (CMB) radiation is the most fundamental limitation on interstellar communication.
In fact, we will find that it is the \emph{only} limit,
in the sense that all other impairments (by which we mean non-idealities like noise and distortion
and Doppler shift) can either be
circumvented by signal design or are introduced by our technology (like receiver-induced noise). 

\doctable
	{notation}
	{Symbols}
	{l p{7cm}}
	{
	\bfseries Variable & \bfseries Meaning 
	\\
	\hline
	$a$ & Relative acceleration of transmitter and receiver in meters/$\text{sec}^2$
	\\
	$\bw$ & Total bandwidth of the signal
	\\
	$B$ & Bandwidth of an energy bundle $h(t)$ in Hz
	\\
	$\bcode$ 
	& Bandwidth of a code word in Hz; equal to the total signal bandwidth $\bw$
	\\
	$\bcoherence$ & Maximum coherence bandwidth in Hz as determined by ISM and motion impairments
	\\
	$\vec c$ & Representation of a codeword as an $M$-dimensional Euclidean vector
	\\
	$D$ & Distance from scattering screen to receiver in meters
	\\
	$\energy$ & Energy in Joules of a single energy burst
	\\
	$\ebit$ & Average energy in Joules required for each information bit
	radiated from the transmit antenna or measured at input to receiver baseband
	\\
	$\ecode$ & Average energy in Joules for each codeword
	\\
	$\bwexpansion$ 
	& Dimensionless bandwidth expansion factor, 
	equal to total signal bandwidth $\bcode$ relative to the information rate $\rate$
	\\
	$f_c$ & Carrier frequency in Hz, equal to the lowest frequency in the passband signal
	\\
	$h(t)$ & Unit-energy waveform used as a building block for constructing codewords; 
	time and frequency translations
	serve as a set of orthonormal functions for representing codewords
	\\
	$\lambda_c$ & Wavelength in meters corresponding to carrier frequency $f_c$
	\\
	$M$ & Dimensionality or degrees of freedom in each codeword waveform
	\\
	$\npsd$ & Power spectral density of the AWGN measured at input to receiver baseband
	in Watts per Hz (or Joules)
	\\
	$\vec N$ & $M$-dimensional vector of independent Gaussian random variables
	\\
	$\ncode$ & Number of codewords in a codebook
	\\
	$\pe$ & Probability of an error in a single information bit
	\\
	$\power$ & Average signal power in Watts measured at input to receiver baseband
	\\
	$\rate$ & Information rate in bits per second
	\\
	$S$ & Dimensionless multiplicative Gaussian random variable representing scintillation
	\\
	$\mathit{SNR}$ & Dimensionless signal-to-noise ratio at input to receiver baseband
	\\
	$T$ & Time duration of a basis function $h(t)$ in seconds
	\\
	$\tau (f)$ & Group delay in seconds as a function of frequency $f$
	\\
	$\Delta \tau$ & Delay spread in seconds, equal to the variation in group delay across the bandwidth
	\\
	$\tcode$ & Time duration of one codeword in seconds
	\\
	$\tcoherence$ & Maximum coherence time in seconds as determined by ISM and motion impairments
	\\
	$x$ & Lateral distance on the scattering screen
	\\
	$z(t)$ & Signal plus noise at the output of a matched filter; $\big| z(0) \big|^2$ is an estimate of the energy of a burst
	\\
	\hline
	}
 
 \begin{changea}
 CMB and other thermal noise sources are modeled by
additive white Gaussian noise (AWGN).\footnote{
''Additive'' means it is added to the signal (as opposed to scintillation
which is multiplicative, see Sec. \ref{sec:scintillation}).
''White'' means that its power spectrum is flat with frequency,
which is true of black body and other forms of thermal radiation
(as long as we stay in the microwave frequencies).
''Gaussian" means that its amplitude has a Gaussian probability distribution.}
For the moment simplify the problem by assuming that CMB is the only impairment,
and ask the question ''what is the best that communication can be''?
Any limit on communication takes into account three elements:
the information rate $\rate$,
the reliability with which the information is communicated
(typically measured by bit error rate $\pe$),
and the resources consumed.\footnote{
A quantity of information is measured by the number of bits required to
represent it.
The units of $\rate$ is thus bits per second,
and also for this reason base two is used in the logarithm in \eqref{eq:shannonRate}.
}
\begin{changec}
In communications there are two limiting resources, the total
bandwidth $\bw$ occupied by the signal
and the “size” of the signal (usually quantified by its average power $\power$).
\end{changec}

AWGN is the simplest case for calculating the Shannon limit.
What this will reveal is a fundamental tradeoff:
For fixed $\rate$ and reliability, lower $\power$ can be
achieved when $\bw$ is larger.
Thus there is an opportunity to substantially reduce the energy
requirements for interstellar communication, if a commensurate increase in $\bw$
can be tolerated.
At the same time, an increase in $\bw$ can simplify the coding and signal processing considerably,
which is consistent with implicit coordination.
This simplification extends to other ISM and motion impairments; see Sec. \ref{eq:limitInterstellar}.
\end{changea}

\subsection{Tradeoff of bandwidth vs power}

The tradeoff between $\bw$ and $\power$ is quantified by the Shannon limit.
\begin{changea}
For this purpose, all the quantities should be measured at the same point
in Fig. \ref{fig:blockDiagram}, the baseband input to the receive signal processing.
In particular $\power$ and $\bw$ are measured at this point, and
the AWGN is assumed to have power spectral density $\npsd$ Watts per Hz.
\end{changea}
Reliable communication, meaning an
arbitrarily low bit error rate $\pe$, is feasible if and only if
\citerefthree{164}{658}{172}
\begin{align}
\label{eq:shannonRate}
&\rate < \bw \cdot \log_{\,2} \left( 1 + \mathit{SNR} \right) \\
\notag
&\mathit{SNR} = \frac{\power}{\npsd \, \bw} \,.
\end{align}
The quantity $\mathit{SNR}$ is interpreted physically as the signal-to-noise power ratio
at the baseband input, since $\npsd \, \bw$ is the total noise power within the signal bandwidth.

\begin{changea}
Many readers may assume that lower $\mathit{SNR}$ is bad, and therefore
$\bw$ should be made as small as possible.
In terms of the achievable $\rate$, \eqref{eq:shannonRate} actually establishes the opposite!
Although the $\log$ term does decrease as $\bw$ increases, the multiplicative
factor $\bw$ in \eqref{eq:shannonRate} more than makes up for this, 
resulting in an overall increase in the feasible $\rate$.
Intuitively this is because increasing $\bw$ admits more degrees of freedom per unit time for
representing information, 
and this beneficial effect dominates any adverse reliability issues due to an increase in total noise.

The bandwidth-vs-rate tradeoff can be highlighted by reworking \eqref{eq:shannonRate} in terms of metrics of
more direct interest to interstellar communication,
\begin{align}
\label{eq:ShannonLimit}
&\frac{\ebit}{\npsd} > \bwexpansion \cdot \left( 2^{1/\bwexpansion} - 1 \right) \\
\notag
&\ebit = \frac{\power}{\rate} \,, \ \  \bwexpansion = \frac{\bw}{\rate} \,.
\end{align}
The metrics $\ebit$ and $\bwexpansion$ are rate-independent.
The energy per bit $\ebit$ is the minimum energy required at the baseband input
for each bit of information reliably communicated.
This is easily related to metrics of interest in the radio portion of Fig. \ref{fig:blockDiagram},
such as the transmit and receive antenna gains and the radiated energy at the transmitter per bit of information.
The bandwidth expansion factor $\bwexpansion$ is the total signal bandwidth
(at either baseband or passband) in relation to the information rate.

This feasible region of \eqref{eq:ShannonLimit} is illustrated as the shaded region Fig. \ref{fig:feasibleEnergyRegion}
\begin{changeb}
over a range of seven orders of magnitude.\footnote{
\begin{changea}
In communications the decibel (dB) is used to represent power or energy radios,
but here (as in Fig. \ref{fig:feasibleEnergyRegion}) orders of magnitude are used
(one order of magnitude equals 10 dB).
\end{changea}
}
\end{changeb}
Smaller bandwidth results from moving to the left, and smaller energy results from moving down.
If the goal is efficiency, or to use the fundamental limit as an implicit form of coordination
as in Fig. \ref{fig:approachingFL},
then an operating point near the boundary of the region will be sought.
The major distinction between interstellar and terrestrial communications
is a difference in the desired location on that boundary.
Terrestrial communication normally operates in the upper left part of the region
($\bwexpansion \ll 1$), where
minimization of bandwidth is prioritized at the expense of greater $\ebit$.
This is because there is an artificial scarcity of spectrum created by regulatory authorities, 
who divide the spectrum among various uses,
and radiated power is rarely a significant cost factor for the short transmission distances involved.

\incfig
	{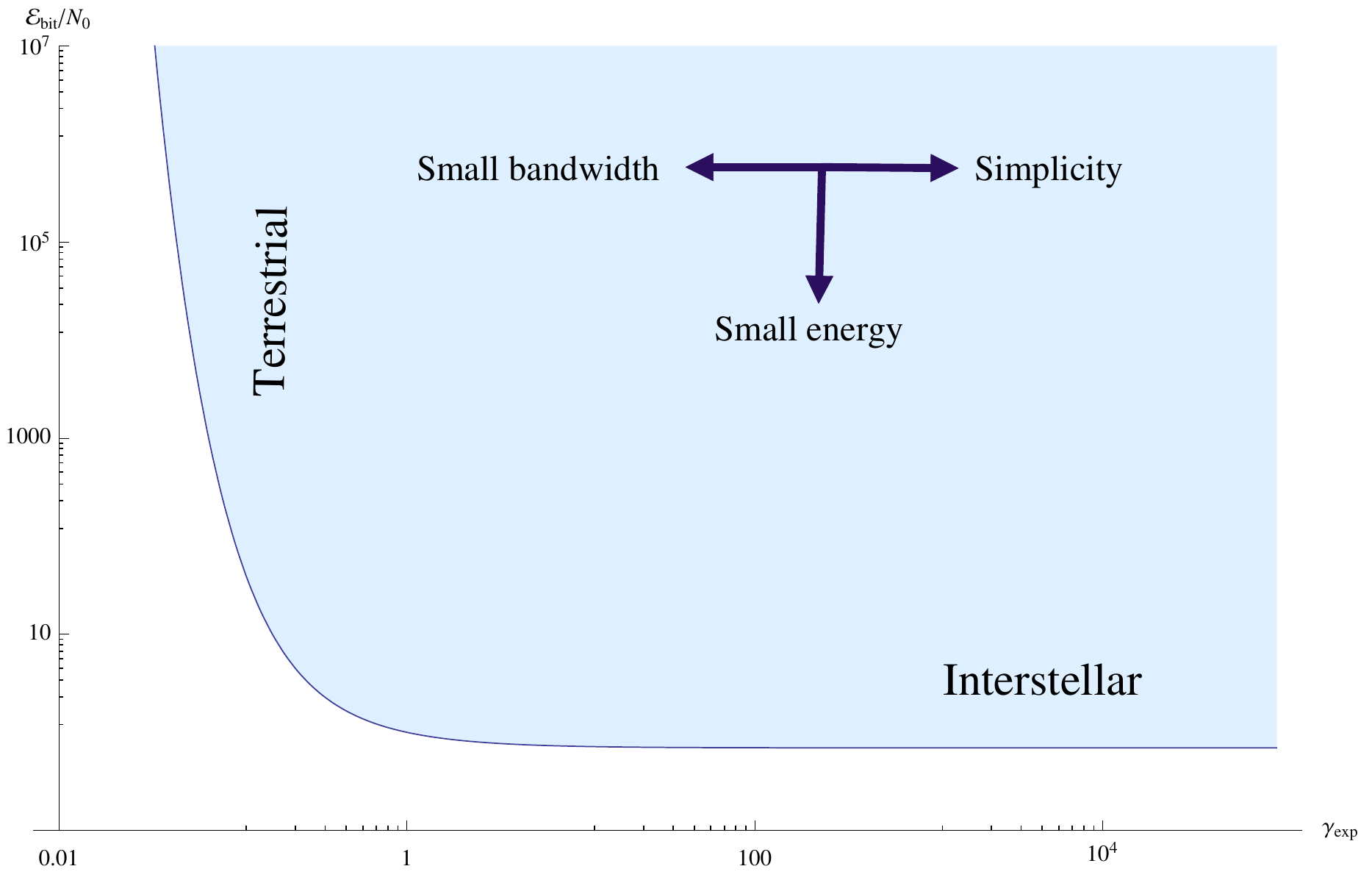}
	{1}
	{A log-log plot of $\ebit/\npsd$ for AWGN
	against the bandwidth expansion $\bwexpansion$.
	Within the shaded region, reliable communication 
	is feasible according to \eqref{eq:ShannonLimit}.}

In interstellar communication, a major source of costs is
energy consumption and the construction of a high-power transmitter
and large-area transmit and receive antennas.
These costs directly trade off against one another, but however that
tradeoff is worked the overall cost is reduced as the delivered energy
to the receiver is reduced.
This argues for the smallest $\ebit$, or
operation in the lower part of the feasible region ($\bwexpansion > 1$).
\end{changea}
In 1995, Jones observed that there is a large window of microwave frequencies 
over which the ISM and atmosphere are relatively transparent \citeref{196}.
Why not, he asked, make use of this wide bandwidth, assuming there are other benefits to be gained?  
Fridman followed up with a specific end-to-end characterization of the energy requirements \citeref{602}.
Neither author takes the ISM and motion impairments into account; see Sec. \ref{sec:ich}.

\begin{changec}
There is an additional motivation for operating in the lower right of Fig. \ref{fig:feasibleEnergyRegion}.
Relatively small $\ebit$ can be achieved with moderate increases in $\bwexpansion$,
but is also consistent with moving to the right toward much larger values of $\bwexpansion$.
This observation is significant because large $\bwexpansion$ also enables major simplifications to the design, 
a crucial contributor to implicit coordination.
This statement is further justified in Secs. \ref{sec:channelCoding} and \ref{sec:ich}.
\end{changec}

\subsection{Minimum delivered energy}

When any bandwidth constraint is removed ($\bwexpansion \to \infty$), \eqref{eq:ShannonLimit} becomes
\begin{equation}
\label{eq:limit}
\frac{\ebit}{\npsd} >  \log 2  = 0.69 \,,
\end{equation}
representing the globally smallest achievable $\ebit$.
The energy penalty for aggressively constraining bandwidth is substantial.

\begin{example}
Some terrestrial standards operate at a bandwidth expansion as low as $\bwexpansion =$ 5\%,
which increases $\ebit$ relative to \eqref{eq:limit} by a factor of $4.6 \times 10^4$.
\end{example}
If minimizing $\ebit$ is a priority, $\bwexpansion > 1$ is mandatory, although
it is not necessary for $\bwexpansion$ to be too large to extract significant energy reductions.
\begin{example}
in \eqref{eq:ShannonLimit}
 $\ebit$ is a factor of $1.44$ above \eqref{eq:limit}
at $\bwexpansion = 1$ and $1.03$ at $\bwexpansion = 10$.
\end{example}
\begin{changec}
The motivation for using $\bwexpansion \gg 1$ is the significant simplification in
signal structure as well as the complexity of the CSP, as discussed in Sec. \ref{sec:channelCoding}.
\end{changec}

Taking only CMB into account, \eqref{eq:limit} predicts that
$\ebit = 8$ photons per bit is required at the receiver at a frequency of 5 GHz.
The transmit energy and power to achieve this requires assumptions as to antenna gains and distance.
\begin{example}
Typical parameters for a starship and a civilization are illustrated in Table \ref{tbl:energyExample}
assuming an ideal implementation.\footnote{
The system parameters are a 5 GHz carrier frequency with circular aperture antennas. 
Everything is assumed to be ideal, including 100\% efficiency antennas, and the only source of noise is the CMB at 2.7 degrees Kelvin.}
Both cases require $\ebit =$ 0.46 Watt-hours at the transmitter, because the chosen transmit antenna
apertures and distances exactly offset.
Taking into account a gap to the fundamental limit as well as nonidealities in the transmitter and receiver
(such as coding and antenna inefficiency and receiver-induced noise), the reality will be on the order of $10$ to $10^2$
greater transmit $\ebit$.
\end{example}
The minimum energy delivered to the receiver for a message consisting of $L$ bits is
$L \, \ebit$, and is independent of the information rate $\rate$.
\begin{example}
A transmit $\ebit$ of 0.46 Watt-hours corresponds to
3.7 megawatt-hours per megabyte. 
This is a substantial energy requirement for a starship, but even on Earth
at typical 
\begin{changeb}
electricity
\end{changeb}
prices this would cost roughly \$400 per megabyte
(\$4000 to \$40,000 per megabyte in practice).
\end{example}
For transmission to a civilization, the energy and cost per message 
is multiplied by repeated transmission of the message on multiple lines of sight simultaneously, 
allowing that the transmitter may not know in advance where the message will be monitored.

\doctable
	{energyExample}
	{Examples of energy requirements at the fundamental limit}
	{lccc}
	{
	\bfseries Parameter & \bfseries Starship  & \bfseries Civilization & \bfseries Units
	\\
	\hline
	Tx antenna diameter & 3 & 300 & meters
	\\
	Rec antenna diameter & 300 & 300 &  meters
	\\
	Distance & 10 & 1000 & light years
	\\
	Received $\ebit$ & 8 & 8 & photons
	\\
	Transmitted $\ebit$ & 0.46 & 0.46 & Watt-hours
	\\
	\hline
	}

As $\rate$ is increased, the message transmission time decreases
at the expense of larger average power $\power$ and a proportionally larger annualized energy cost.
Particularly for communication with a civilization
as argued in \citeref{602}
it should be acceptable to use low information rates
in the interest of reducing annualized costs.
\begin{example}
At $\rate = 1$ b/s the annual transmitted information is 4 megabytes,
for the parameters of Table \ref{tbl:energyExample}
consuming 15 megawatt-hours of energy at the fundamental limit, and costing 
at current Earth energy prices about \$1600
(in practice closer to \$16K to \$160K).
This assumes that the signals are designed to
minimize energy (including a large $\bwexpansion$) as described in the sequel.
\end{example}
It is significant that
a low information rate makes a large $\bwexpansion$ more palatable.
\begin{changec}
This is fortuitous, since as shown in Secs. \ref{sec:channelCoding} 
and \ref{sec:ich} larger values of $\bwexpansion$ yield significant simplifications.
\end{changec}
\begin{example}
For $\rate = 1$ b/s, $\bwexpansion = 10^6$ requires only a megahertz, 
which is tiny when compared to the available microwave window, and implementation of
 $\bwexpansion = 10^9$ would be straightforward with current Earth technology.
\end{example}

\section{Channel coding}
\label{sec:channelCoding}

Although \eqref{eq:ShannonLimit} mandates an expansion in bandwidth in order to reduce energy,
it offers no insight into why this is the case.
\begin{changea}
To understand this, it is useful to examine how energy can be reduced in practice
by designing an appropriate \emph{channel code}, 
one of the core functions in the coding and signal processing
block of Fig. \ref{fig:blockDiagram}.
\begin{changeb}
The channel code function maps information bits into
a baseband continuous-time waveform suitable for passing to the
radio subsystem.
Its design determines the actual operating point 
in Fig. \ref{fig:feasibleEnergyRegion}, and thus can impact the
performance parameters over many orders of magnitude.
\end{changeb}
See \citerefthree{172}{658}{43} for further details on channel coding.

\subsection{Morse's code and the energy burst}

\begin{changeb}
Simplicity is critical in the absence of coordination, so it is informative to examine the
very first example of a channel code, that invented in 1836
by Samuel Morse for the telegraph \citeref{667}.
This code maps one bit of information into an \emph{carrier burst} waveform.
A typical burst waveform $h(t)$ at baseband is shown 
on the left side of FIg. \ref{fig:sinusoidalEnBunWaveform}.
Although Morse used a square pulse (implemented as opening a closing a switch),
to be somewhat bandwidth efficient a similar smooth waveform is shown.
After translation to passband, this waveform would be a burst of sinusoid as illustrated on the right side
(this is the continuous-wave or CW code still widely used by amateur radio operators).
Building on the carrier burst, Morse mapped one bit of information into
a ''dot'' and a ''dash'', where the duration of the ''dash'' is longer than the ''dot''
(and hence its energy is correspondingly greater).

The essential idea behind Morse's code applied to radio is to represent information by the energy
in a waveform, meaning
there is no information represented by the phase or the amplitude of the passband sinusoid.
We adopt the more general term \emph{energy burst} for this building block,
recognizing that a sinusoid is only one possible realization \citeref{211}.
When digital (as opposed to analog) radio communication was revived more than a century following Morse,
it was noted that increasing bits of information could be represented
by transmitting a set of distinct phases or amplitudes
without a significant increase in bandwidth
(since changing the phase or the amplitude of a sinusoid has no impact on its bandwidth).
Modern terrestrial radio or wireless systems use complicated renditions
of these schemes \citeref{574}, which are ideal for achieving a small $\bwexpansion$.
Although appropriate for the bandwidth-scarce terrestrial context,
this is an inappropriate direction for interstellar communication
because it inevitably results in a larger $\ebit$.
A multiplicity of phases or amplitudes are more easily confused by noise,
requiring an increase in $\ebit$ to achieve equivalent reliability.

\incfig
	{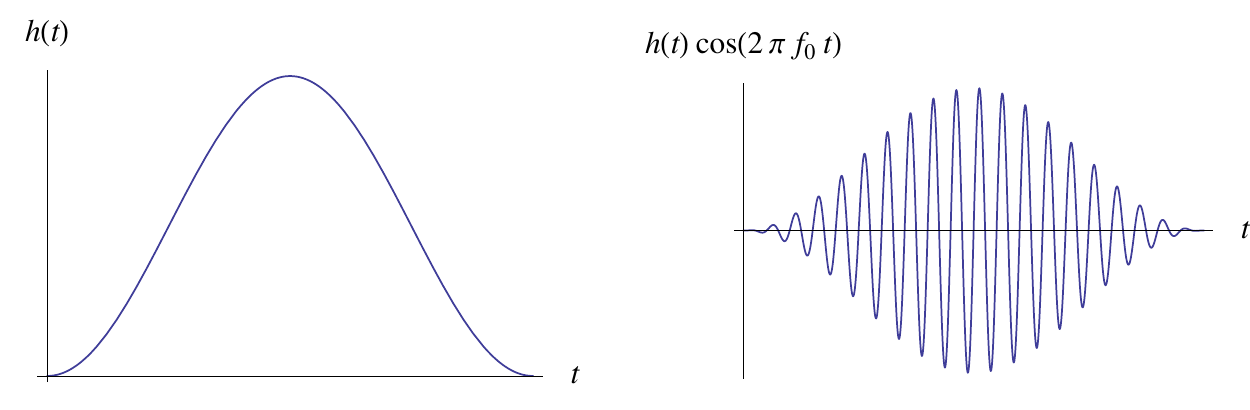}
	{1}
	{
	An energy burst is represented at baseband as a continuous waveform $h(t)$
	as shown on the left.
	This raised cosine function is
	designed to minimize bandwidth and sidelobes.
	On the right is a typical waveform when $h(t)$ is shifted in frequency
	for purposes of modulation to passband or as needed for the location code of Fig. \ref{fig:centauriDreams}.
	}

The minimum $\ebit$ is always achieved (at the expense of a larger $\bwexpansion$), like Morse,
by mapping a single bit onto a waveform.
An example is on-off keying, where one bit is mapped onto a burst or no burst ($h(t)$ or zero),
and this already achieves a relatively low $\ebit$.
Is there a direction in which on-off keying can be modified
so as to reduce $\ebit$ further, and even
(if we are lucky) approach the fundamental limit of \eqref{eq:limit}
at the expense of increasing bandwidth?

The answer to this question is yes, and the simplest such method known represents \emph{multiple bits}
of information by the location of a \emph{single} energy burst in time and frequency.
We adopt the name \emph{location coding}, and it represents one special case of the \emph{orthogonal coding}
proposed in \citeref{196}.
Specifically it represents multiple bits of information by the location of a single
energy burst in time and/or frequency.
Using location (as opposed to phase or amplitude) increases the
signal bandwidth, but holds the energy relatively fixed as the number of
bits represented increases.
(Quantifying that price in energy requires detailed consideration of the
reliability implications, which comes later.)
Thus, location coding is the ''dual'' of phase/amplitude coding,
modifying Morse's idea in a direction of energy minimization as opposed to bandwidth minimization.
\end{changeb}

\subsection{Location coding}

\begin{changeb}
Location coding is illustrated by the scheme shown in Fig. \ref{fig:centauriDreams}.
This coding represents multiple bits of information by the location of a single burst of energy
in a two-dimensional grid of times and frequencies.
Each codeword (shown as a dashed rectangle) has a large number of locations,
only one of which contains an energy burst (shown as a black dot).
Each energy burst consists of a waveform $\sqrt{\energy} \,h(t)$ translated in time and frequency,
and the receiver estimates the value of the energy $\energy$ for each location.
Having thus determined which location contains the burst, it can infer the information bits.
\end{changeb}
This consumes extra bandwidth to accommodate many locations,
only one of which is in actual use within any one codeword.
This is consistent with \eqref{eq:ShannonLimit}, 
in that extra bandwidth is necessary to achieve energy efficiency.

\incfig
	{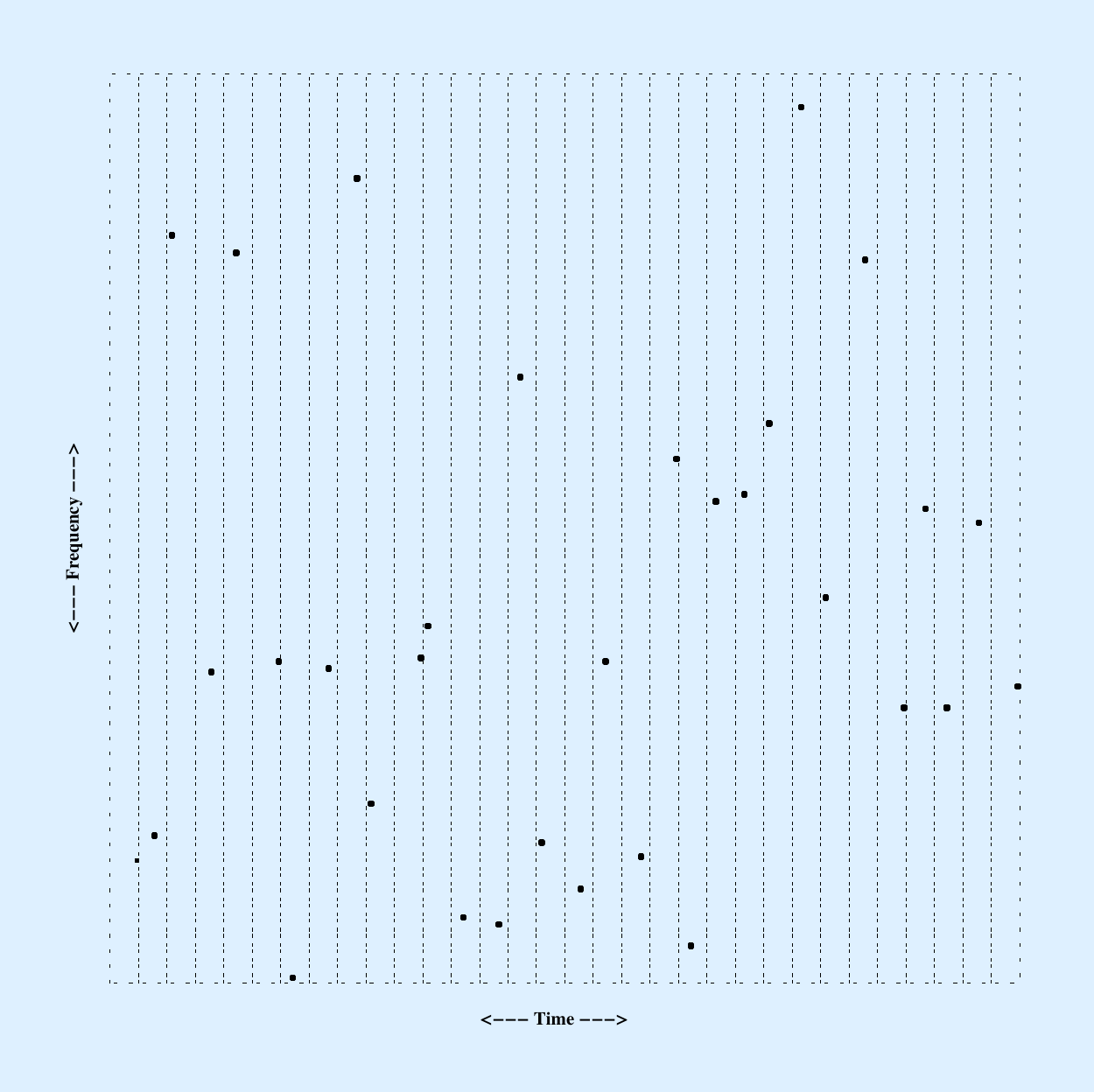}
	{1}
	{
	An example of an information-bearing signal based on block location coding.
	The dashed rectangles represent codewords, each containing $2048$ possible locations.
	Only a single location is used for an energy burst, and is represented by a black dot.
	}

\begin{example}
\begin{changeb}
Each codeword (dashed rectangle) in Fig. \ref{fig:centauriDreams} represents a codeword
conveying 11 bits of information to the receiver by using one of $2^{11} = 2048$ possible locations.
Those locations are spaced on a grid with 8 times (representing 3 bits) and 256 frequencies (representing 8 bits).
\end{changeb}
If $\rate = 1$ b/sec, then each codeword has a time duration of $11$ seconds,
which is divided into 8 timeslots each of duration $11/8 = 1.375$ seconds.
A single energy burst requires a minimum bandwidth about equal to the reciprocal of its time duration,
which is $0.73$ Hz. The total bandwidth is thus about 
$0.73 \times 256 = 186.2$ Hz, and thus $\bwexpansion \approx 186.2$.
\end{example}

The motivation for using a time-bandwidth grid of locations is to permit
the signal to circumvent all ISM and motion impairments 
(with the exception of noise and scintillation).
We show in Sec. \ref{sec:ich} that the energy burst waveform $h(t)$ for each location in this grid
avoids these impairments if its time duration and bandwidth is appropriately restricted.
Transmitter and receiver processing then focuses on noise and scintillation, requiring absolutely no processing
related to other ISM and motion impairments.

In light of its simplicity,
it is remarkable that a location code of this type can actually
achieve the lowest possible $\ebit$ asymptotically as the duration of its codewords
and its bandwidth increases.
\begin{changec}
Intuitively this is because one constraint (bandwidth) is removed, simplifying the design.
\end{changec}
The code structure also
follows from a principled design approach that takes into account the characteristics of
the ISM and motion impairments.
To appreciate this, we have to delve further into channel code design.
\end{changea}

\subsection{A primer on channel coding}

A coding structure such as the location code of Fig. \ref{fig:centauriDreams}
is actually the culmination of a principled design process.
Delving into this more deeply
leads to insights into what types of channel codes should be used with
starships and with civilizations, and the signal structure to expect.

\subsubsection{Codebooks}

A \emph{block} channel code has the structure illustrated in Fig. \ref{fig:codewordStructure},
which uses a sequence of codewords, each of duration $\tcode$.
Assume the information rate is $\rate$.
Then over a time duration $\tcode$, the number of bits of information that must
be transmitted is $\rate \, \tcode$.
In a channel code, this is accomplished by transmitting one of $\ncode$ waveforms
 distinguishable at the receiver, where
\begin{equation}
\label{eq:ncode}
\ncode = 2^{\, \rate \, \tcode} \,.
\end{equation}
Each such waveform is called a \emph{codeword}, 
and the \emph{codebook} is the set of $\ncode$ distinct codewords.
At the receiver, the received signal for a duration $\tcode$ is examined to determine 
(subject to errors caused by noise and other impairments) 
which codeword was transmitted, with the result used to recover the $\rate \, \tcode$ bits.

\incfig
	{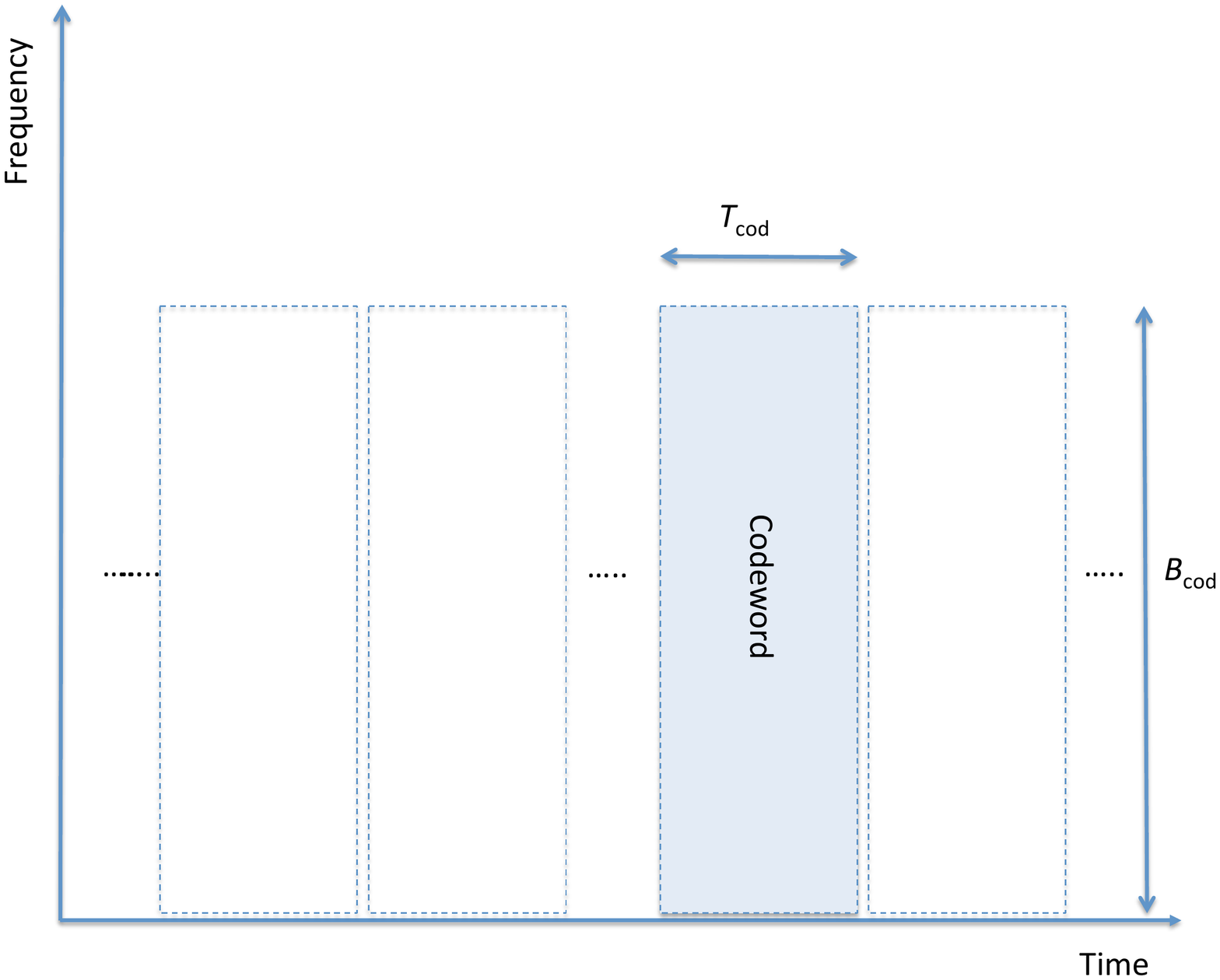}
	{1}
	{
	A channel code associates a set of information bits with a codeword,
	which is transmitted as some waveform occupying a time duration $\tcode$
	and bandwidth $\bcode$.
	}

\begin{changea}
\begin{example}
If $\rate = 1$ b/s with the channel code of Fig. \ref{fig:centauriDreams},
the parameters are $\tcode = 11$ s, and $\ncode = 2^{1 \times 11} = 2048$
There are $2048$ codewords in the codebook, and
each codeword consists of an energy burst in a 
distinct location with $2048$ possibilities.
\end{example}
\end{changea}
	
\subsubsection{Extending time duration of a codeword}

Suppose that an energy $\ecode$ is allocated to each codeword on average
(it is possible for different codewords to have different energy, but the average is what matters).
An average signal power $\power$ can be assured if
\begin{equation}
\label{eq:ecode}
\ecode = \power \, \tcode \,.
\end{equation}
As $\tcode \to \infty$, both $\ncode \to \infty$ and $\ecode \to \infty$, and
\begin{equation*}
\ebit = \frac{\power}{\rate} = \frac{\ecode}{\log_2 \ncode} \,.
\end{equation*}
Based on \eqref{eq:ncode} and \eqref{eq:ecode}, $\ebit$ remains fixed as $\tcode \to \infty$.
For AWGN, there exists a sequence of codebooks such that
the average probability of error in a single bit $\pe \to 0$ as $\tcode \to \infty$ 
if and only if \eqref{eq:limit} is satisfied.

\subsubsection{Basis functions}
\label{sec:basis}

\begin{changea}
Each codeword is a continuous-time waveform, but since its time duration $\tcode$ and
bandwidth $\bcode$ are finite it actually has finite degrees of freedom.
For example, the sampling theorem establishes that it
can be represented in terms of $M = \bcode \tcode$ complex-valued samples
at rate $\bcode$.
It turns out that the  $\mathrm{sinc} (\cdot )$ functions used in the sampling
theorem are inappropriate in this case because they suffer from
dispersion in the ISM, so consider the more general representation
\begin{equation*}
c(t) = \sum_{k=1}^{M} c_k \, f_k (t) \,.
\end{equation*}
This divides the codebook design issue into
choosing a set of $M$ orthonormal basis functions
$\{ f_k (t) \,, \ 1 \le k \le M \}$ and
a set of $M$ coordinates $\{ c_k \,, \ 1 \le k \le M \}$ with respect to those basis functions.
In the sequel it is assumed that basis waveforms are chosen
to be impervious to impairments introduced by the ISM and motion.
This implies that the integrity of codewords is maintained at the receiver,
and also provides valuable guidance in the choice of basis functions.
\end{changea}
 
\subsubsection{Detection of codewords}

\begin{changea}
Once a set of basis functions is chosen, it is convenient to
consider the coordinates of the codewords to be $M$-dimensional Euclidean vectors.
\end{changea}
The codebook can be represented by a set of $\ncode$ such vectors,
\begin{equation}
\label{eq:codebook}
\trans{\vec c_i = [ c_{1,i} , \, c_{2,i}, \, \dots , \, c_{M,i}]} , \ \  1 \le i \le \ncode\,,
\end{equation}
where $\trans{\vec x}$ denotes the transpose of $\vec x$.
\begin{changea}
When the received baseband signal is represented in terms of the same basis functions,
\end{changea}
it becomes
\begin{equation}
\label{eq:receptionVector}
\vec y =\vec c_m + \vec N
\end{equation}
where $m$ is the index of the transmitted codeword,
and $\vec N$ contains $M$ statistically independent complex-valued circular Gaussian noise samples, 
each with variance $\expec{|N_i |^2} = \npsd$.

\begin{changea}
An analysis of
the detection problem for \eqref{eq:receptionVector} finds that the smallest $\pe$ 
is obtained by picking the codeword closest in Euclidean distance to $y$.
\end{changea}
The resulting reliability depends on $\npsd$ and the
Euclidean distance $|| \vec c_i - \vec c_j ||$ between codewords.
In particular, $\pe$ is dominated by the
codeword pairs that are closest and thus most easily mistaken due to the additive noise.

\subsubsection{Codebook design}

The design challenge can be stated as finding a codebook consisting of a set of $\ncode$ codewords
as in \eqref{eq:codebook}.
This choice is subject to a constraint on the average energy (assuming codewords are equally likely)
\begin{equation}
\label{eq:energyconstraint}
\frac{1}{\ncode} \sum_{i=1}^{\ncode} || \vec c_i ||^{\,2} = \ecode \,,
\end{equation}
and attempts to achieve the largest minimum Euclidean distance between pairs of codewords.

\subsubsection{Matched filter}
\label{sec:MFdesc}

\begin{changeb}
In the presence of AWGN, there is an optimum way to estimate the energy $\energy$ of a burst.
That is a cross-correlation of signal plus noise with the complex-conjugate of burst waveform $h(t)$.
The signal component of this cross-correlation equals the energy $\energy$,
\begin{equation*}
\left|\,
\int_{0}^{T} \sqrt{\energy} \, h(t) \cdot \cc h(t) \, \diff t \, \right|^2 = 
\energy \int_{0}^{T} \big| h(t) \big|^2 \, \diff t = \energy \,.
\end{equation*}
In practice $h(t)$ may have an unknown delay, so calculation of the cross-correlation
is desired for all possible lags.
This is accomplished by the \emph{matched filter}, which is a filter with impulse response
$\cc h( - t )$ and frequency response $\big| H(f) \big|^2$.

All the following results presume matched filtering, which requires knowledge of $h(t)$
(see \citeref{211} for a discussion of this).
There are two possible sources of sensitivity degradation: (a) any mismatch between the
$h(t)$ assumed by transmitter and receiver and (b) any changes to $h(t)$ as it
propagates through the ISM and due to motion effects.
\end{changeb}

\subsubsection{Interstellar coherence hole}
\label{sec:ichprelim}

As discussed in Sec. \ref{sec:ich}, 
the combination of impairments from the ISM and from motion results in
a characteristic coherence time $\tcoherence$ and a coherence bandwidth $\bcoherence$
for a particular line of sight and carrier frequency.
The coherence time captures an interval short enough that the end-to-end propagation
can be accurately approximated as time-invariant, and the bandwidth
captures the range of frequencies small enough that any variation in magnitude or phase
is sufficiently small to be neglected.

Consider the transmission of a single waveform $h(t)$ through the ISM.
If $h(t)$ has time duration
$0 \le T \le \tcoherence$ and bandwidth $0 \le B \le \bcoherence$,
then $h(t)$ is said to fall in the \emph{interstellar coherence hole} (ICH).
\begin{changea}
The terminology ''hole'' captures the opportunity for a waveform $h(t)$
to propagate through the ISM essentially absent any impairment,
\end{changea}
\begin{changeb}
specifically as measured by an undesirable reduction in the estimate of energy appearing at the output of a
filter matched to $h(t)$.
\end{changeb}
Any distortion of $h(t)$ is small enough to be neglected,
eliminating any and all processing in the transmitter and
receiver related to impairments and eliminating whole dimensions of search during discovery.

\subsubsection{ICH orthogonal basis}

\begin{changea}
The ICH provides specific guidance on the choice of basis functions in Sec. \ref{sec:basis},
because if each basis function falls in the ICH then each codeword is unaffected
by transmission through the ICH and by motion impairments.
The most obvious basis, the $\mathrm{sinc} (\cdot )$ functions of the sampling theorem,
would inevitably violate the frequency coherence constraints of the ICH as $\bcode$ grows.
A suitable basis can be formed from any waveform $h(t)$ falling in the ICH.
Given such an $h(t)$ with time duration $T$ and bandwidth $B$, the set of functions
\begin{equation*}
f_{m,k} (t) = e^{\,i\, 2 \pi m B t} h(t - k \, T) \,.
\end{equation*}
for integer values of $m$ and $k$ fall in the ICH and
are mutually orthogonal because there is no overlap in time and/or frequency.
\end{changea}
Choosing these as basis functions leads to the codeword structure
illustrated in Fig. \ref{fig:codewordInternalStructure}.
\begin{changea}
The coordinates of a codeword correspond to dividing the entire time-bandwidth
product into a two-dimensional grid.
\end{changea}

	\incfig
	{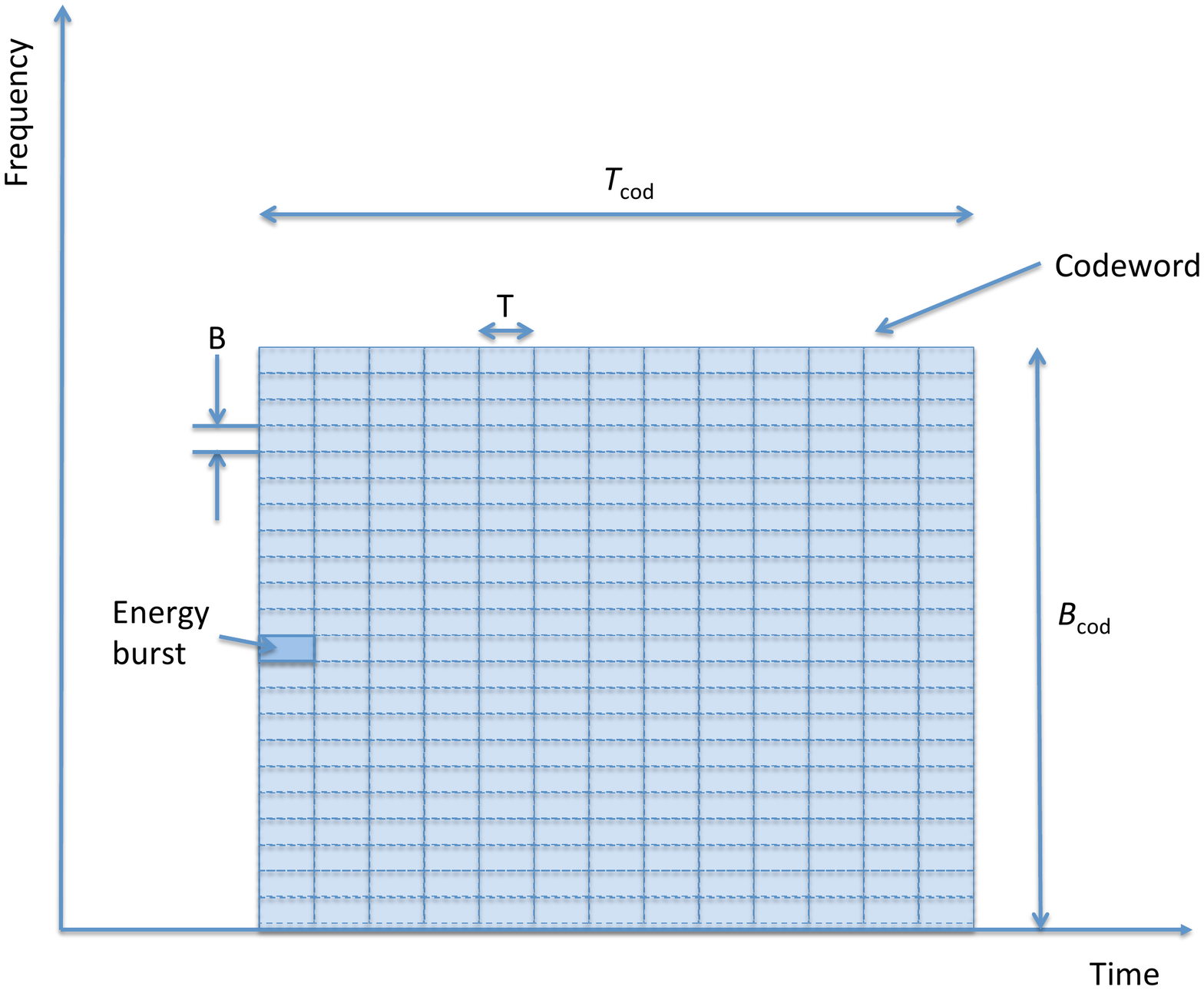}
	{1}
	{
	An illustration of an orthogonal basis for the codeword shown in Fig. \ref{fig:codewordStructure},
	chosen such that each basis function falls in the ICH.
	}

While a $\mathrm{sinc} (\cdot)$ waveform has time-bandwidth product $B \, T \approx 1$,
$h(t)$ can have $B \, T \gg 1$, for example to provide greater immunity to
radio-frequency interference (RFI) \citeref{211}.
After the choice of $B \, T$,
the total number of basis functions in total codeword time duration $\tcode$ and total 
codeword and signal bandwidth $\bcode$ is
\begin{equation}
\label{eq:mBcodeTcode}
M = \frac{\bcode \, \tcode}{B \, T} \,.
\end{equation}
While the largest $M$ for a given value of $\bcode$ occurs with the minimum value $B\,T \approx 1$,
RFI would suggest choosing $B\,T \approx \bcoherence \, \tcoherence$
and increasing $\bcode$ to compensate.
The receiver is advised to search for both of these obvious choices.

Fig. \ref{fig:codewordsOrthogonal} illustrates typical
codebooks, where each codeword is represented by a Euclidean vector as in \eqref{eq:codebook}.
As $\tcode \to \infty$, a major distinction that
captures the essence of the tradeoff between bandwidth and energy
develops between unconstrained and constrained bandwidth.

\subsection{Unconstrained bandwidth}
\label{sec:unconstainedBW}

The ideal codebook has the largest possible minimum distance between codewords.
There is greater freedom to manipulate this minimum distance when the bandwidth
(or equivalently the dimensionality $M$) is unconstrained.

The maximum possible distance between two codewords $\vec c_i$ and $\vec c_j$
is predicted by the triangle inequality
\begin{equation*}
|| \vec c_i - \vec c_j || \le || \vec c_i || + || \vec c_j || \,,
\end{equation*}
but this bound is achievable only if the two codewords are collinear.
For example, when $||c_i|| = ||c_j|| = \sqrt{\energy}$ then $\vec c_j = - \vec c_i$ achieves
$|| \vec c_i - \vec c_j || = 2 \sqrt{\energy}$.
However this most favorable case does not extend beyond two codewords.

The alternative of orthogonal
codewords such as in the location codebook
($\hc{\vec c_i} \vec c_j = 0$
where $\hc{\vec x}$ is the conjugate transpose of $\vec x$)
is almost as good, since by the Pythagorean theorem
\begin{equation*}
|| \vec c_i - \vec c_j || = \sqrt{\, || \vec c_i ||^2 + || \vec c_j ||^2 } \,,
\end{equation*}
and for equal-length codewords
($|| \vec c_i || = || \vec c_j || = \sqrt{\energy}$)
the distance between codewords
$|| \vec c_i - \vec c_j || = \sqrt{2 \,\energy}$
is only a factor of $\sqrt 2$ smaller.
In return for this slightly smaller distance,
as many as $M$ codewords can be orthogonal, and further
each and every pair of codewords achieves this same distance.
If $M$ is unconstrained, this means an arbitrary number of
codewords can be orthogonal and benefit from this favorable spacing.
An orthogonal codebook is illustrated in $M=3$ dimensions
in Fig. \ref{fig:codewordsOrthogonal}.

\incfig
	{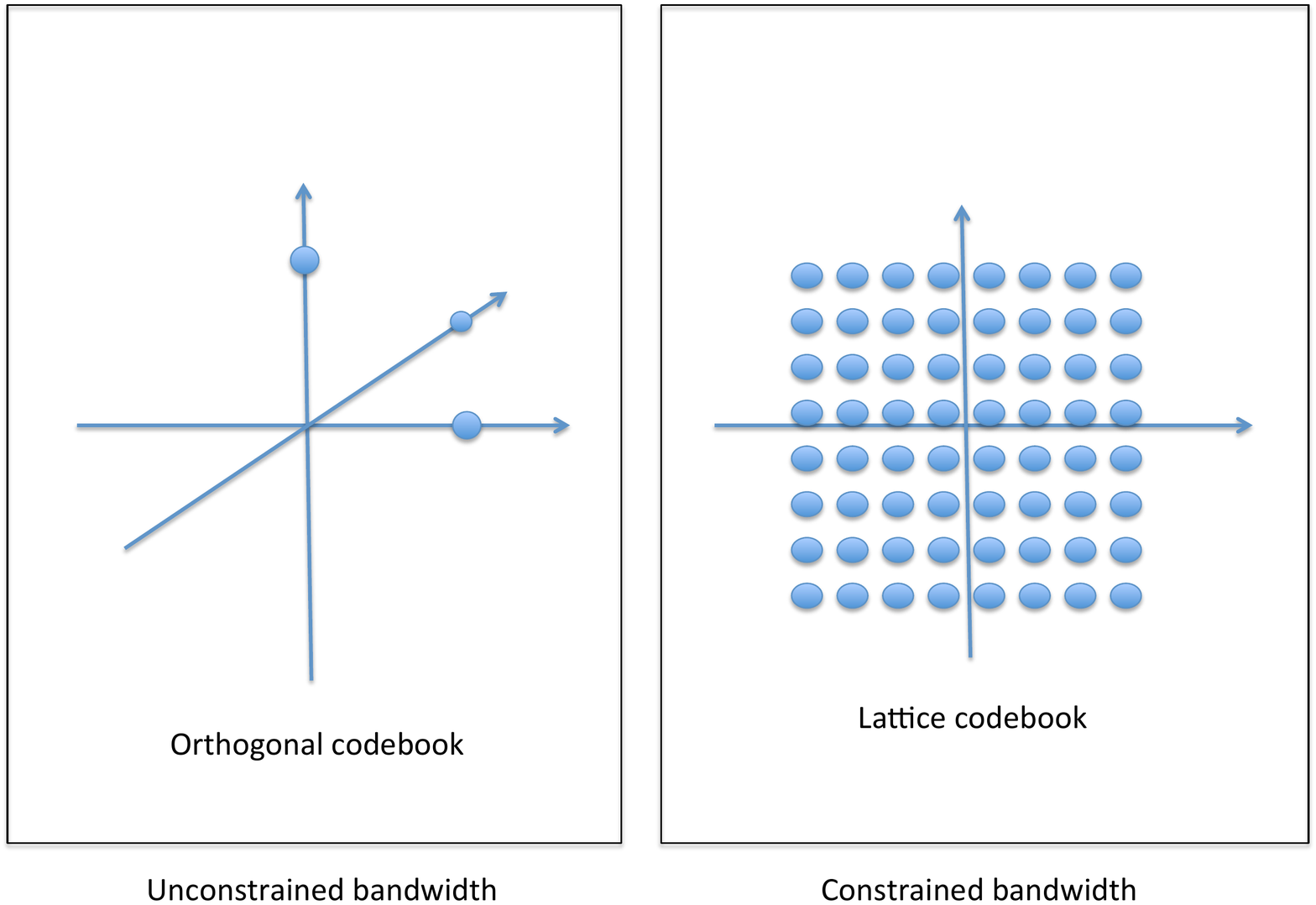}
	{.8}
	{
	A comparison between an orthogonal codebook in three dimensions and a lattice codebook in two dimensions.
	When bandwidth is unconstrained, another codeword can be added to an orthogonal codebook without
	reducing the spacing between codewords. 
	When bandwidth is constrained the number of dimensions is constrained, and adding codewords
	reduces the spacing. 
	 }

The simplest orthogonal codebook chooses $M = \ncode$,
the minimum permissible dimensionality permitting $\ncode$ orthogonal codewords.
In terms of minimum distance, there is no advantage to choosing a larger $M$.
It then concentrates all the energy
of each codeword in a single basis waveform
so all coordinates are zero except the $i$-th,
\begin{equation}
\label{eq:cartesian}
\vec c_i = \trans{[0, \, 0 \, \dots , \, \sqrt{\ecode} , \, \dots \ \, 0 , \, 0]} \,.
\end{equation}
The resulting bandwidth expansion
\begin{equation}
\label{eq:bwExpOrthogonal}
\bwexpansion = \frac{\bcode}{\rate} =  B \, T \cdot \frac{2^{\, \rate \, \tcode}}{\rate \, \tcode}
\end{equation}
grows without bound when $\tcode \to \infty$.

\begin{changea}
\begin{example}
The location code of Fig. \ref{fig:centauriDreams} is a special case of orthogonal codewords.
The energy burst waveforms corresponding to different locations do not overlap
in time or frequency, and are thus orthogonal.
With minimum burst bandwidth $B \, T = 1$ the bandwidth expansion for this code is 
$\bwexpansion = 2^{11}/11 = 186.2$.
If $\tcode$ were to be doubled, then $\bwexpansion = 2^{22}/22 = 1.9 \times 10^5$.
Each codeword would have $16 = 2^4$ timeslots and $2^{18} = 262144$ frequencies.
\end{example}
\end{changea}

It was observed in the 1950's \citeref{639} that
$\pe \to 0$ as $M \to \infty$ if and only if
\eqref{eq:limit} is satisfied
(see \citerefWloc{658}{Sec.8.5} for a modern development).
More recently an orthogonal codebook with finite-but-large $M$ has been proposed for interstellar 
communication because of its energy efficiency properties \citereftwo{196}{602}.

Fig. \ref{fig:centauriDreams} illustrates what happens when a
location codebook \eqref{eq:cartesian} is combined with the time-frequency basis of
Fig. \ref{fig:codewordInternalStructure}.
The singular characteristic of this signal is energy bursts isolated in 
discrete but sparse locations in time and frequency. 
\begin{changea}
Each codeword consists of a single energy burst conveying the entire codeword energy $\ecode$,
with the codewords distinguished only by the location of the bursts.
\end{changea}
Each burst has to be sufficiently energetic to overwhelm the noise at the receiver, so that its location can be detected reliably. 

This is  how a lighthouse works: Discrete flashes of light are each energetic enough to overcome loss and noise, but they are sparse in time (in any one direction) to conserve energy. 
This is also how optical SETI is usually conceived, 
because optical designers are usually unconcerned with bandwidth \citeref{655}.
Cost considerations in high-power radio transmitter design also suggests using a low duty factor \citeref{568}.
There is, however, one major distinction between these examples and the energy burst in
Fig. \ref{fig:centauriDreams}.
The waveform conveying an individual energy burst 
should be consciously chosen to fall in the ICH 
 as described in Secs. \ref{sec:ichprelim} and \ref{sec:ich}, and matched filter detection has to be used at the receiver
 if the lowest delivered energy is to be achieved.
 
 There are numerous design opportunities for a high-power transmitter generating a signal such as
 Fig. \ref{fig:centauriDreams}, including parallelism in frequency and sequentially scanning multiple targets
 using an antenna array without increasing peak radiated power.

Approaching the fundamental limit of \eqref{eq:limit} always depends on allowing $\tcode \to \infty$.
For the codebook of \eqref{eq:cartesian}, there are two competing factors working to determine the reliability.
From \eqref{eq:ecode}, the energy $\ecode$  of the single energy burst
increases in proportion to $\tcode$, improving noise immunity.
However, the number of false codewords at distance $\sqrt{2 \, \ecode}$ from the actual codeword
equals $(\ncode-1)$, which from \eqref{eq:ncode} increases exponentially with $\tcode$.
The first factor dominates the second, allowing $\pe \to 0$, but only when \eqref{eq:limit} is satisfied.
Illustrating this,
$\ebit$ vs $\bwexpansion$ is shown in
Fig. \ref{fig:orthogonalEbN0} for a fixed value of $\pe = 10^{\,-3}$ and various values of $M$.
$\ebit$ asymptotically approaches twice the fundamental limit of \eqref{eq:limit}.\footnote{
Specifically this is a plot of \eqref{eq:peOrthogonal}, which also assumes phase-incoherent
detection as described in Sec. \ref{sec:phaseIncoherentDetection} 
and hence always suffers a factor of two penalty relative to \eqref{eq:limit}.
This penalty can be avoided by adding time diversity as described in Sec. \ref{sec:timeDiversity}.
}

Signals resembling Fig. \ref{fig:centauriDreams} have a very different character from those customary in terrestrial radio communication. 
This is an advantage in itself because another big challenge not yet mentioned is confusion with artificial signals of terrestrial or near-space origin. 
Confusion is less of a problem if the signals (local and interstellar) are distinctly different.

\subsection{Constrained bandwidth}

\begin{changea}
If bandwidth is considered a resource that comes at a considerable cost,
then $\bwexpansion$ can be reduced without a sacrifice in $\ebit$.
The $\ebit$ achieved by the location code is shown in Fig. \ref{fig:orthogonalEbN0},
together with the fundamental limit from \eqref{eq:ShannonLimit}.
A location codebook can achieve only $\bwexpansion \ge 2$, so
the plot is restricted to this region.
This codebook does consume voluminous bandwidth in 
the interest of low energy, since it achieves the lowest energy only
asymptotically as $\bwexpansion \to \infty$.
It is feasible to operate in the shaded region, with 
a more favorable tradeoff between $\bwexpansion$ and $\ebit$.
This can be explored by constraining $\bwexpansion$, and
designing codebooks that drive down $\ebit$.
However, this pays a price in complexity as now explained,
and may be difficult to achieve in the absence of coordination.
\end{changea}

\incfig
	{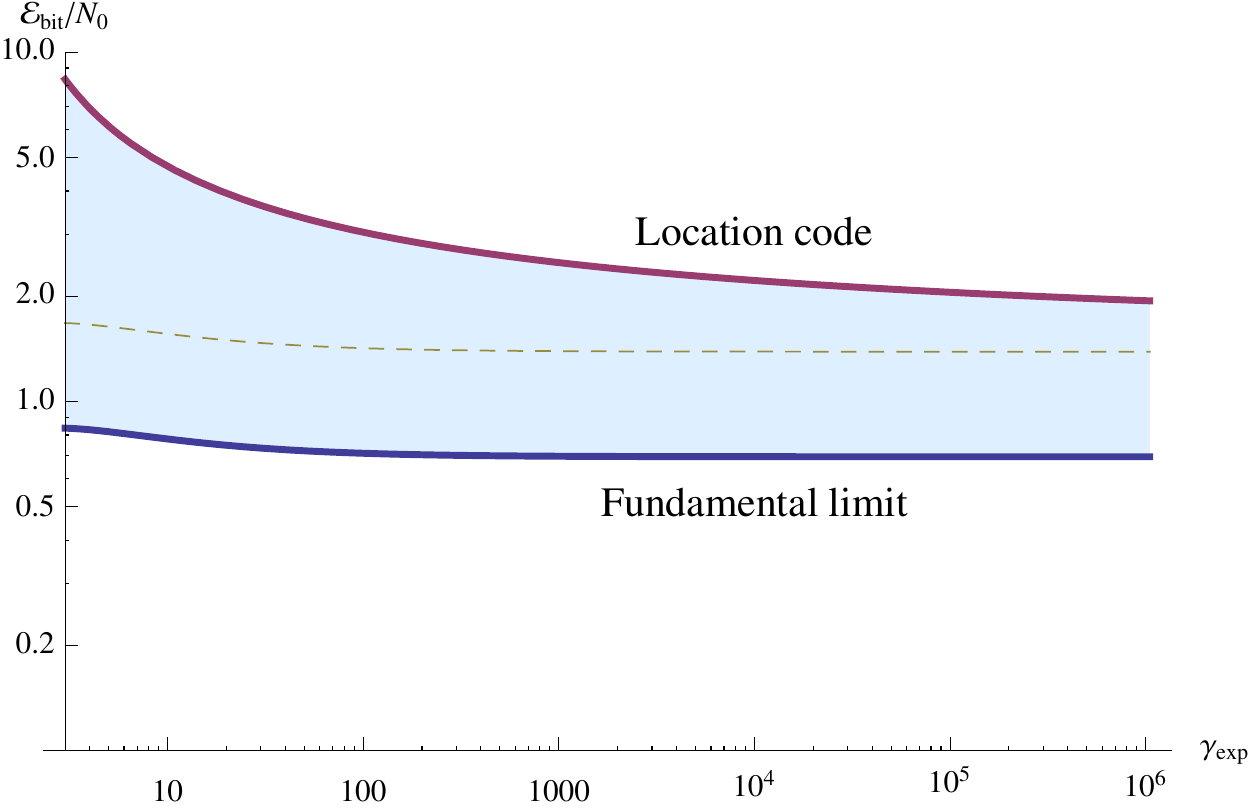}
	{1}
	{
	A log-log plot of the energy per bit $\ebit/\npsd$ vs the dimensionality of each
	codeword $M$ for a location code, at a bit error rate
	 of $\pe = 10^{\,-3}$ obtained from \eqref{eq:peOrthogonal}.
	Also shown is the fundamental limit of \eqref{eq:ShannonLimit}
	for the same bandwidth expansion $\bwexpansion$ as the location code
	and (dashed line) twice that fundamental limit (which accounts for phase-incoherent detection).
	The shaded area shows the region where more complex codes can in principle achieve both a lower $\ebit$
	and greater reliability than the location code for a given $\bwexpansion$.
	}

What happens when $\bw$ (and hence $\bcode$) is constrained?
From \eqref{eq:ncode} $\ncode$ increases exponentially with $\tcode$,
but \eqref{eq:mBcodeTcode} permits only a linear increase in $M$ with $\tcode$.
Thus as $\tcode \to \infty$ eventually $\ncode \gg M$ is inevitable, and there
are insufficient dimensions to use mutually orthogonal codewords exclusively.
Thus, a situation pictured in Fig. \ref{fig:codewordsOrthogonal} is forced,
where codewords cannot all be orthogonal and must be packed into a lower dimensional space.
When this results in a reduction in minimum distance,
noise immunity suffers.
Unconstrained bandwidth allows codewords to
have a higher dimensionality, in effect providing more space to
keep codewords farther apart.

Constraining bandwidth is certainly reasonable for communication with starships,
since the design is fully coordinated.
A question is whether constrained bandwidth is feasible for communication with
a civilization, absent any coordination.
While the lattice packing shown in Fig. \ref{fig:codewordsOrthogonal} looks simple, in fact
designing codebooks in higher dimensionality (for large $\tcode$) is a very challenging
problem due to the exponential explosion in complexity as $\tcode$ increases.
It took decades of research to even arrive at feasible solutions in terrestrial systems \citereftwo{618}{651},
and those solutions are arguably too complex to be considered in the absence of coordination.
Nevertheless further research is appropriate, seeking simpler constrained-bandwidth cookbooks
that can reasonably be reverse engineered at the receiver
armed with nothing more than observation of the signal.

\section{Interstellar coherence hole}
\label{sec:ich}

The impairments introduced by the ISM and motion are now analyzed, with
an emphasis on mechanisms that limit the time and frequency coherence.
While the physical mechanisms
at work in the ISM and motion are quite different from terrestrial wireless,
the signal impairments that result are familiar and thus the terrestrial experience is 
directly applicable to the communication design problem \citeref{574}.

A set of orthogonal basis waveforms is the first step in the design of a codebook.
Achieving the minimum energy delivered to the receiver requires matched filtering
to each codeword, which in turn requires matched filtering to individual basis waveform $h(t)$.
\begin{changea}
A filter matched to $h(t)$ forms the correlation of the noisy baseband reception 
with a time-delayed and conjugated
$\cc h (t-\tau)$ for different values of $\tau$, thus determining the presence and location of $h(t)$.
\end{changea}
In practice there will always be unknown parameters for $h(t)$, such as
time duration and bandwidth and time scaling, and this will require searching over
these parameters in the discovery phase \citeref{211}.
The more that can be inferred or guessed about the characteristics of $h(t)$ in advance,
the smaller the resources required in a discovery search, 
the lower the false alarm probability, and the greater the receiver sensitivity.
The specific question now addressed is the degree to which the interstellar impairments jointly
observable by transmitter and receiver constrain the parameterization of $h(t)$.

For earthbound observers, pulsar astronomy provides observations
and models pertinent to the choice of time duration and bandwidth for $h(t)$ \citeref{212}.
\begin{changea}
In particular, pulsar observations and other physical considerations
have reverse-engineered the ISM impairments.
Based on the ISM and an understanding of motion impairments as well,
the maximum duration
(called the coherence time $\tcoherence$) and 
a maximum bandwidth (called the coherence bandwidth $\bcoherence$) 
can be estimated (with some remaining uncertainty) for a particular line of sight and distance.
Together these maximums define the ICH
\end{changea}
and inform the choice of $h(t)$ in the transmitter
as well as dramatically narrow the scope of a discovery search.
This does not suggest a specific waveform for $h(t)$, although other
principles can be invoked \citeref{211}.

This section focuses on physical arguments for the existence of the ICH,
and also conveys an intuitive sense of how the size of the ICH can be estimated,
with some details relegated to Appendix \ref{sec:ismModel}.
Refining the accuracy of these estimates requires more detailed modeling \citeref{656}.

\subsection{Energy burst detection}

Discovery requires a search over both carrier frequency $f_c$
and starting time.
For each such value,
the received waveform referenced to baseband is assumed to be $h(t)$,
a complex-valued signal (Appendix \ref{sec:signal}).
The in-phase and quadrature carrier signals at passband correspond to
the real and imaginary parts of $h(t)$ respectively.
Assume that $h(t)$ is confined to time $0 \le t \le T$,
has Fourier transform $H(f)$ confined to frequency band $0 \le f \le B$, and 
has unit energy.

\subsubsection{Matched filter}

Following demodulation, the resulting baseband signal $y(t)$ is applied to a
filter matched to $h(t)$ (see Sec. \ref{sec:MFdesc}).
\begin{changea}
This matched filter is conveniently represented as a convolution
(with symbol $\convolve$),
\end{changea}
\begin{equation}
\label{eq:MF}
z(t) = y(t) \convolve \cc h(t) = \int_{0}^{T} \cc h(u) \, y(t-u) \, \diff u \,,
\end{equation}
where $\cc h(t)$ is the conjugate of $h(t)$.
This matched filter is the optimal front-end processing
for countering AWGN, but does not take direct account of other impairments.
When the signal component of $y(t)$ is $\sqrt{\energy} \, h(t)$,
where $\energy$ is the signal energy at baseband,
then the matched filter output sampled at time $t = 0$
($\Re \{ z(0) \}$, where $\Re \{\cdot\}$ denotes real part) is a noise-corrupted estimate of $\sqrt{\energy}$.
\begin{changea}
Neither the time duration $T$ nor the bandwidth $B$ of $h(t)$ affect the sensitivity or the error probability;
only energy matters.
\end{changea}

\subsubsection{Phase-incoherent detection}
\label{sec:phaseIncoherentDetection}

Consider an unknown phase $\theta$, assumed fixed over the time duration of $h(t)$.
This phase is due to an unknown transmitted carrier phase and imprecise knowledge of propagation distance.
Then $\sqrt{\energy} \, h(t)$ is replaced by $e^{\,i\,\theta} \sqrt{\energy} \, h(t)$, 
and the signal portion of $\Re \{ z(0) \}$
becomes $\Re \{ e^{\,i\,\theta} \, \sqrt{\energy} \} = \sqrt{\energy} \, \cos ( \theta )$,
depending strongly on $\theta$.
A simple way to counter this is to use $|z(0)|^2$ as an estimate of $\energy$,
which is then called \emph{phase-incoherent} matched filtering.
This introduces a factor of two noise penalty, since the estimate is corrupted by
both the imaginary and real parts of the noise.

\subsection{Coherence hole definition}

Other interstellar impairments introduce more subtle distortions of $h(t)$.
Suppose the result is an actual received waveform $g(t)$ rather than $h(t)$,
also with unit energy.
Then the energy estimate becomes
\begin{equation*}
|z(0)|^2 = \energy \, \big| \, g(t) \convolve \cc h(t)  \, \big|^2_{t=0} \le \energy \,,
\end{equation*}
with equality if and only if $g(t) = h(t)$ by
the Schwartz inequality.
Any impairments causing $g(t) \ne h(t)$ have the effect of reducing the energy estimate.
The definition of the ICH is a coherence time $\tcoherence$ and coherence bandwidth $\bcoherence$
such that if $T \le \tcoherence$ and $B \le \bcoherence$, then
$|z(0)|^2 \approx \energy$ with whatever accuracy is demanded.
\begin{changea}
When $h(t)$ violates the constraints of the ICH,
the energy estimate $|z(0)|^2 \ll \energy$ and receiver sensitivity is impacted.
\end{changea}

Two classes of impairments illustrated in Fig. \ref{fig:ICH-illustration}
determine the size of the ICH, caused by four distinct physical phenomena.
Impairments due to the relative motion of transmitter and receiver cause
time-varying effects and time incoherence, and
impairments due to propagation through the ISM
cause a time-invariant dispersion or spreading of $h(t)$.
In particular,
propagation of radio waves over galactic distances is affected by
turbulent clouds of gasses that are conductive due to partial ionization of these gasses
and the resulting free electrons, constituting a low-temperature and low-density plasma \citeref{598}.

\incfig
	{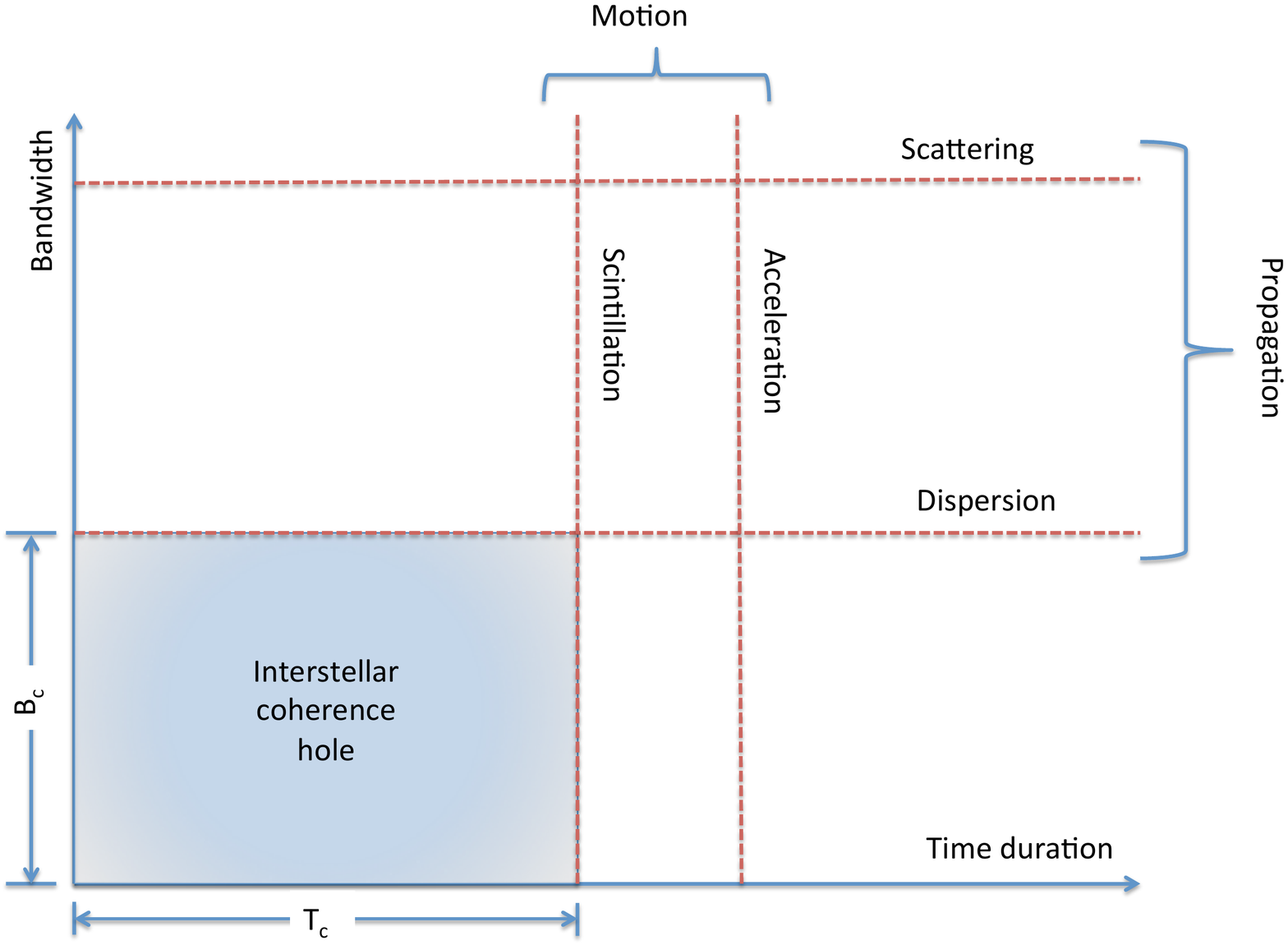}
	{1}
	{Illustration of the interstellar channel impairments affecting the size
	of the interstellar coherence hole (ICH),
	including the coherence time $\tcoherence$ and coherence bandwidth $\bcoherence$.
	$\tcoherence$ is determined by the more stringent of scintillation 
	and acceleration.
	$\bcoherence$ is determined by the more stringent of dispersion
	and scattering.}

When $h(t)$ is chosen by the transmitter to fall in the ICH and
any residual distortion is neglected,
the resulting decision variable $Q = | z (0) |^2$ at the matched filter output
can be modeled as
\begin{equation}
\label{eq:channelModel}
Q = \big|\, \sqrt{\energy} \, S + N \, \big|^2 \,,
\end{equation}
where $N$ is a complex-valued Gaussian random variable, and 
$S$ is a multiplicative noise called \emph{scintillation}.
 The only germane signal parameter is the received energy $\energy$,
and notably irrelevant are the time duration $T$, bandwidth $B$, or waveform of $h(t)$.
Additive noise $N$ is attributable to
the CMB as well as other sources of thermal noise such as star noise
and noise introduced in the receiver.
$S$ is attributable to scattering
(see Sec. \ref{sec:scattering}) and will vary from one time to another (see Sec. \ref{sec:scintillation}).
$S$ is the one impairment attributable to the ISM that remains 
when $h(t)$ is confined to the ICH.
For starships and nearby civilizations, the relatively short distances
will result in $|S| \equiv 1$ with random phase, 
but as distance increases $S$ evolves into
a zero-mean complex-valued circular Gaussian random variable
with uniformly distributed random phase and 
Rayleigh-distributed amplitude and unit variance ($\expec{| S |^2} = 1$) \citeref{656}.

\subsection{Frequency coherence}
\label{sec:freqCoherence}

If the conditions for time coherence of Sec. \ref{sec:timeCoherence}
are satisfied, then time-varying phenomena can be neglected.
In that case, aside from the multiplicative $S$ factor,
the remaining dispersive impairments can be modeled by a
frequency response $G(f) = |G(f)| \, e^{\,i\,\phi(f)}$.
Two related dispersive phenomena have been observed by
pulsar astronomers: plasma dispersion and scattering.
Both are illustrated by a simplified one-dimensional model in Fig. \ref{fig:scatteringIllus}.
The free electrons in the ISM turbulent gas clouds create a conductive 
medium which causes a phase shift which can vary with both wavelength and the spatial dimension.
The question is how small bandwidth $\bcoherence$ must be so that the
frequency dependence of $G(f)$ over
$f_c \le f \le f_c + \bcoherence$ is sufficiently small to be neglected.

\incfig
	{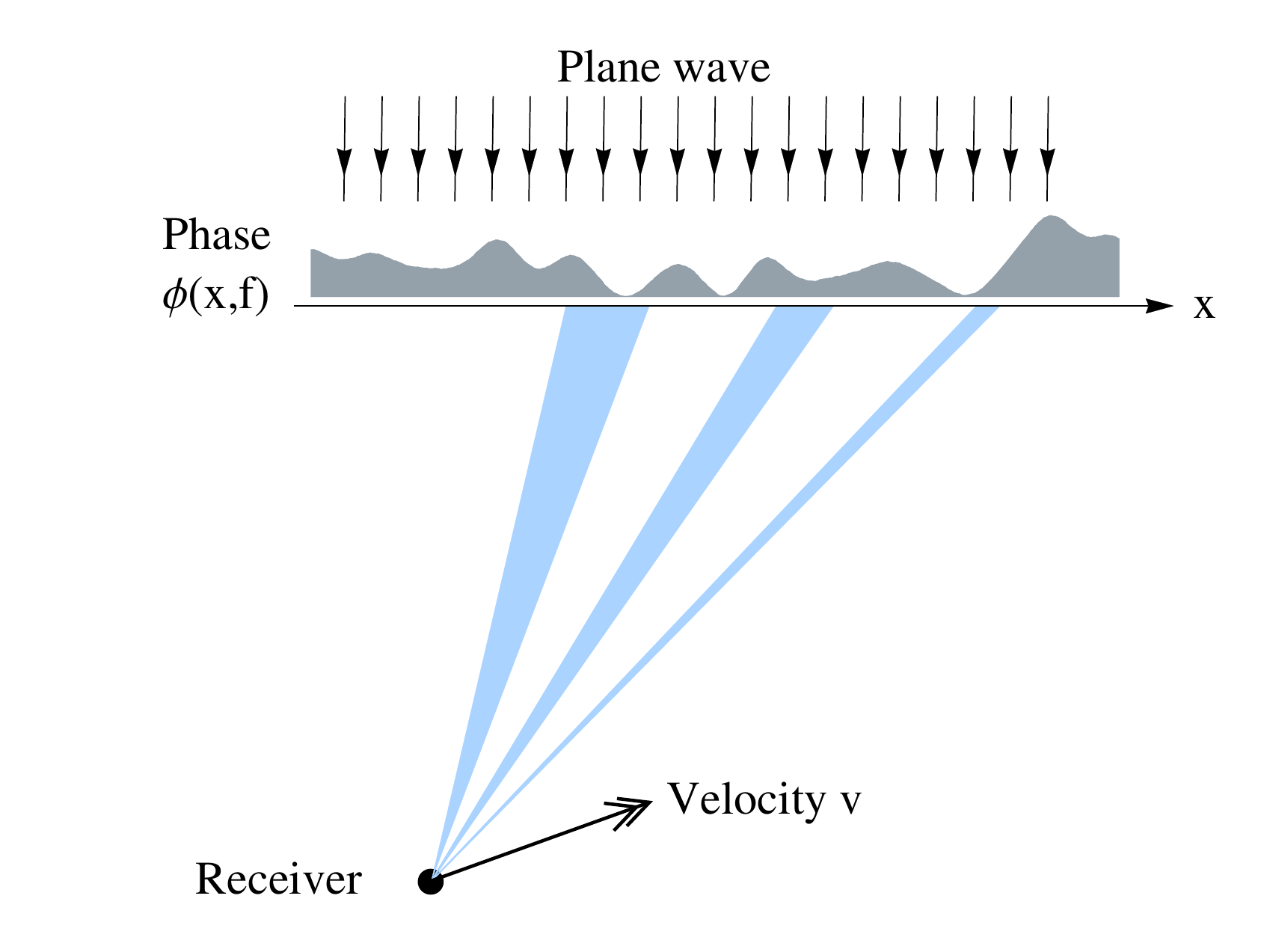}
	{.8}
	{
	In a simplified model, inhomogeneous clouds of ionized electrons
	are represented by a variation in phase shift $\phi (x)$ as a function
	of lateral distance $x$ on a one dimensional scattering screen.
	If the source is far away, incoming electromagnetic radiation can be approximated as a
	plane wave arriving at the screen.
	The receiver sees a superposition of
	rays arriving from each position $x$ on the screen.
	The receiver is assumed to
	be in motion with velocity $\vec v$ relative to the line of sight.
	}

\subsubsection{Delay spread}

A convenient way to represent the phase portion of the frequency response
is by the equivalent group delay $\tau (f)$.
Physically $\tau (f)$ captures
the delay experienced by energy at frequency $f$
(see \eqref{eq:groupDelay} for a mathematical definition).
The \emph{delay spread} $\Delta \tau$ is the
variation in group delay over the entire signal bandwidth $0 \le f \le B$.
When the bandwidth is sufficiently small, the signal changes slowly
enough that the effect of group delay is not noticeable.
Specifically the dispersive effect on $h(t)$ is negligible (see Appendix \ref{sec:plasmaDispersionModel})
when the bandwidth is constrained by
\begin{equation}
\label{eq:freqCoherence}
B  \ll  \frac{1}{\Delta \tau} \,.
\end{equation}

\subsubsection{Plasma dispersion}

In plasma dispersion,
free electrons absorb an incident photon and re-emit it with a 
photon energy-dependent group delay.
If propagation through the ISM is confined to a single path
(neglecting the multiple paths due to scattering),
this effect is lossless and fully characterized by its group delay $\tau (f)$.
The standard model used in pulsar astronomy is
\begin{equation*}
\tau (f) = \frac{\alpha}{f^2}
\end{equation*}
where $\alpha$ is an observable parameter that is proportional to the
columnar density of electrons (average number of electrons falling in a cylinder)
along the line of sight \citeref{212}.
It follows that
\begin{equation*}
\Delta \tau = \frac{\alpha}{f_c^{\,2}} - \frac{\alpha}{(f_c+B)^{\,2}} \approx \frac{2 \, \alpha \, B}{f_c^{\,3}} \,,
\end{equation*}
and based on \eqref{eq:freqCoherence}
the coherence bandwidth $B \propto f_c^{\,3/2}$
increases with increasing carrier frequency.
Typical values for $B$ range from about a kHz to a MHz \citeref{656}.

\subsubsection{Scattering}
\label{sec:scattering}

The inhomogeneity in the density of ionized gas clouds 
as illustrated in Fig. \ref{fig:scatteringIllus} causes
a variation in phase with both lateral distance and frequency,
and this in turn is the source of scattering \citereftwo{598}{582}.
These density variations are at an extremely large scale relative to a wavelength,
and thus a ray tracing model (often used in geometrical optics) is accurate.
The energy arriving at the receiver from different rays add destructively or constructively
depending on their relative phase shift.

Consider a patch of the scattering screen $x_0 \le x \le x_0 + \Delta x$.
To result in strong constructive interference, and hence a larger contribution to
energy, two conditions must be satisfied.
First, the variation in phase across the patch due to plasma inhomogeneity must be small.
The largest scale over which this typically occurs (as modeled by the statistics of plasma turbulence)
is called the \emph{diffraction} scale.
Second, the variation in phase due to the geometrical difference in
path length must be small.
The scale over which this occurs is the \emph{geometric} scale.
The interplay of these two scales leads to two distinct regimes:
\begin{itemize}
\item
\emph{Weak} scattering occurs when the diffraction scale is large
relative to the geometric scale, in which case all the energy arriving at the
receiver arrives with essentially the same group delay, and the
main source of dispersion is limited to plasma dispersion.
This is the dominant effect at shorter distances and larger carrier frequencies.
Any scattering encountered in starship communication is likely to be weak.
\item
When the geometric scale is large relative to the diffraction scale,
the scattering is said to be \emph{strong}.
This is characterized by energy arriving with distinct delays corresponding to
different patches of coherent phase on the scattering screen,
the differences in delay corresponding to different propagation distances
(termed \emph{multipath distortion} in communications).
Strong scattering is the dominant effect in communicating with civilizations
at distances of hundreds of light years or more, especially at lower carrier frequencies.
\end{itemize}

Strong scattering can cause a variation in $\big| G(f) \big|$ with $f$, but only if the
bandwidth $B$ is too large.
Consider a superposition of different replicas of waveform $h(t)$
arriving from different locations on the scattering screen,
\begin{align}
\notag
\sum_k r_k \, e^{\,i\, \theta_k} \, h(t - \tau_k )
&\approx \left( \sum_k r_k \, e^{\,i\, \Theta_k} \right) \, h(t) \\
\label{eq:centralLimit}
&= S \cdot h(t)
\end{align}
where the $\{ r_k , \, \theta_k , \, \tau_k \}$ are fixed but unknown amplitudes,
phases, and group delays.
The approximation neglecting  $\{ \tau_k \}$ holds if two conditions are satisfied.
First, the variation in group delay $\Delta \tau$
must be bounded, or in other words there is a largest group delay $\tau_k$
beyond which the $r_k$`s are small enough to be neglected.
It is argued in Appendix \ref{sec:geometry} that this will always be true
based on geometry alone, since increasing $x_0$ (and hence $\tau$)
reduces the geometric scale $\Delta x$, 
and hence less energy incident to the scattering screen has an opportunity to constructively interfere.
Second, the bandwidth $B$ of $h(t)$ must satisfy \eqref{eq:freqCoherence}
for the largest $\tau_k$ of interest.

The effect of scattering on $\bcoherence$ is usually dominated by plasma dispersion \citeref{656}.
Scattering nevertheless does cause an unknown amplitude and phase factor $S$ in \eqref{eq:centralLimit},
the same factor that appears in \eqref{eq:channelModel}.
If the $\{ \theta_k  \}$ are modeled as independent random variables
uniformly distributed on $[0 , \, 2 \pi ]$,
a Central Limit Theorem argument establishes that
$S$ is Gaussian distributed with $\expec{S} = \expec{S^2} = 0$.
A physical argument establishes that $\expec{|S|^2} = 1$,
or in words the \emph{average} signal energy
$\expec{\energy |S|^2}$ is unaffected by scattering
because the scattering screen is lossless \citeref{598}.
There is a stochastic variation in signal energy about that average,
and due to scintillation (Sec. \ref{sec:scintillation}) this variation is
manifested by a difference in the value of $S$ for energy bursts
separated in time or frequency.

\subsection{Time coherence}
\label{sec:timeCoherence}

Inevitably the transmitter and receiver are in relative motion,
resulting in a changing path length from transmitter to receiver
and changes to the scattering geometry.
This can result in time-varying changes to $h(t)$.
However, if the time duration $T$ of $h(t)$ is sufficiently small, those changes
are sufficiently small that they can be neglected.
There are two germane parameters of motion,
acceleration along the line of sight (which results in a time-varying Doppler shift)
and velocity transverse to the line of sight (which results in scintillation).

\subsubsection{Doppler}

The textbook ''frequency shift'' model of Doppler does not apply to a
wideband signal.
The effect is modeled in Appendix \ref{sec:motionModel} with an assumption
 of non-relativistic motion.
Neither the time scale of a signal from a civilization nor its carrier frequency is known.
An unknown relative velocity adds a greater uncertainty to these parameters,
and thus can be neglected.
In starship communication, both the scale and frequency are known,
and relative velocity should be accurately known from navigation information, so
these effects can be compensated at the terrestrial end.
On the other hand, relativistic effects are quite significant for starships
and are deserving of detailed analysis in their own right
\begin{changea}
\citeref{665}.
\end{changea}

Acceleration is more significant than velocity because it causes time incoherence.
It has two deleterious effects on a passband broadband signal.
The first is a phase shift in $h(t)$ that varies quadratically with time, and
can be neglected (Appendix \ref{sec:motionModel}) if 
\begin{equation}
\label{eq:accelphase}
T \ll \sqrt{\frac{\lambda_c}{a}} \,,
\end{equation}
where $a$ is the acceleration and $\lambda_c = c/f_c$ is the wavelength.
The second is a quadratic warping of the time axis of $h(t)$, but this
is insignificant whenever \eqref{eq:accelphase} is satisfied.
Typical values for $T$ in \eqref{eq:accelphase} vary from 0.5 to 10 seconds \citeref{656}.

With a highly directional transmit antenna,
the transmitter and receiver can each correct for the effect of the
acceleration along the line of sight attributable to their motion
within an inertial frame centered on their local star \citeref{662}.
Those two inertial frames may be influenced differentially by
galactic-scale gravitational effects such as dark matter and gravity waves,
but these effects should be small.
Over time frames of interest,
relative star motion should not contribute any appreciable component of
acceleration, and double correction will be effective 
in eliminating these impairments due to acceleration within a small uncertainty.

\subsubsection{Scintillation}
\label{sec:scintillation}

The component  of the transmitter's and receiver's velocity transverse to the
line of sight interact with any multiple propagation paths due to strong scattering.
This is illustrated in Fig. \ref{fig:scatteringIllus} by a vector velocity $\vec v$
for the receiver relative to the scattering screen.
The result is a changing group delay due to changing propagation delay between
any point $x$ on the scattering screen and the receiver.
The group delay is affected slightly differently for different rays, with the component of $\vec v$ 
transverse to the line of sight the dominant contributor.
The resulting change in phase for each ray indirectly changes their
constructive and destructive interference, resulting in a changing value for
$S$ in \eqref{eq:centralLimit} and \eqref{eq:channelModel}.
Since $|S|$ has a Rayleigh distribution, in communications this phenomenon
is called Rayleigh \emph{fading}, while in astronomy it is called \emph{scintillation}.

Changes in $S$ with time due to
scintillation can be neglected for sufficiently small time duration $T$,
with typical coherence times $T$ ranging from $10^2$ to $10^4$ seconds \citeref{656}.
Although $S$ can then be considered constant for one energy burst, it assumes different
values for different energy bursts sufficiently separated in time or frequency.
In \eqref{eq:centralLimit} the phase of $S$ is much more sensitive to receiver
position than is $|S|$, so it is phase variation that dominates any residual incoherence.
Thus, when looking at successive energy bursts, the relative phase shift
in $S$ is randomized, but $|S|$ will be highly correlated.
The value of $S$ will also be different at two different carrier frequencies with spacing
larger than the scattering coherence bandwidth, which varies widely
between 1 kHz and 1 GHz \citeref{656}.

Acceleration dominates the overall coherence time
$\tcoherence$ unless the transmitter and receiver both compensate for its effects,
in which case scintillation dominates.

\subsubsection{Starship communication}

The ISM impairments will be negligible in communication with a starship or a nearby civilization.
As a result, frequency coherence is not an issue,
and the effect of scintillation on time coherence is also not an issue.
Time coherence is thus dominated by relativistic motion effects,
which can possibly be corrected based on starship navigation information.
The general idea of trading large $\bwexpansion$ for reduced $\ebit$ remains valid,
including the use of a location codebook.
However, the underlying basis waveform $h(t)$ can be chosen more freely subject only
to the constraint $B \, T > 1$, and this choice will strongly influence the parameterization of the
time-frequency structure of codewords in Fig. \ref{fig:codewordInternalStructure}.

\section{Fundamental limit: Interstellar communication}
\label{eq:limitInterstellar}

The fundamental limit on received energy in the presence of AWGN
was considered in Sec. \ref{sec:energyLimitGaussian}.
Once impairments due to the ISM and motion have been characterized,
extending that fundamental limit to interstellar communication can be addressed.
The conclusion is that
for the case of an unconstrained bandwidth,
the same energy per bit $\ebit$ as \eqref{eq:limit} can be achieved  \citeref{656}.
With a bandwidth constraint, the modeling is greatly complicated and the
effect of impairments on the achievable $\ebit$ is unknown,
although an increase in the achievable $\ebit$ is to be expected.

\subsection{Scintillation and the limit}

Were it not for scintillation, this fundamental limit would not be surprising since communication can be
based on building a codebook using basis functions which fall in an ICH as illustrated in 
Fig. \ref{fig:codewordInternalStructure}.
Using such basis waveforms that are non-overlapping in time and in frequency is moderately
bandwidth inefficient, but that does not matter if bandwidth is unconstrained.

However, any such scheme must still deal with scintillation $S$ of \eqref{eq:channelModel}
at greater distances, and it is perhaps surprising
that lossless scintillation (for which $\expec{|S|^2} = 1$) 
does not fundamentally and adversely affect the reliability of communication
that can be achieved.
Kennedy first demonstrated in 1964 that
lossless Rayleigh fading in conjunction with Gaussian noise 
does not increase the required $\ebit$  \citeref{637}
(see also \citerefthree{635}{638}{656}).
Intuitively
this is because codewords can be sufficiently spread out in time to average the scintillation over both
favorable and unfavorable periods.
Thus, scintillation is another impairment that can be circumvented by appropriate measures
in transmitter and receiver.

\subsubsection{Time diversity}
\label{sec:timeDiversity}

Proofs of Kennedy's result make use of \emph{time diversity},
which is a specific simple way of spreading out codewords in time.
The starting point is a location codebook that ignores scintillation.
Then as illustrated in Fig. \ref{fig:timeDivIllus},
the total energy $\ecode$ is divided equally among
$J > 1$ replicas of each of these location codewords
spaced at intervals larger than the scintillation coherence time.
At the receiver, the energy estimates for each time-frequency location
are averaged across these replicas to obtain a more reliable estimate of $\ecode$.
The location codebook of Sec. \ref{sec:unconstainedBW} can achieve
reliable communication in the presence of scintillation with
the addition of time diversity as long as \eqref{eq:limit} is satisfied \citeref{656}.
Specifically $\pe \to 0$ as $\tcode \to \infty$ and $J \to \infty$.

\incfig
	{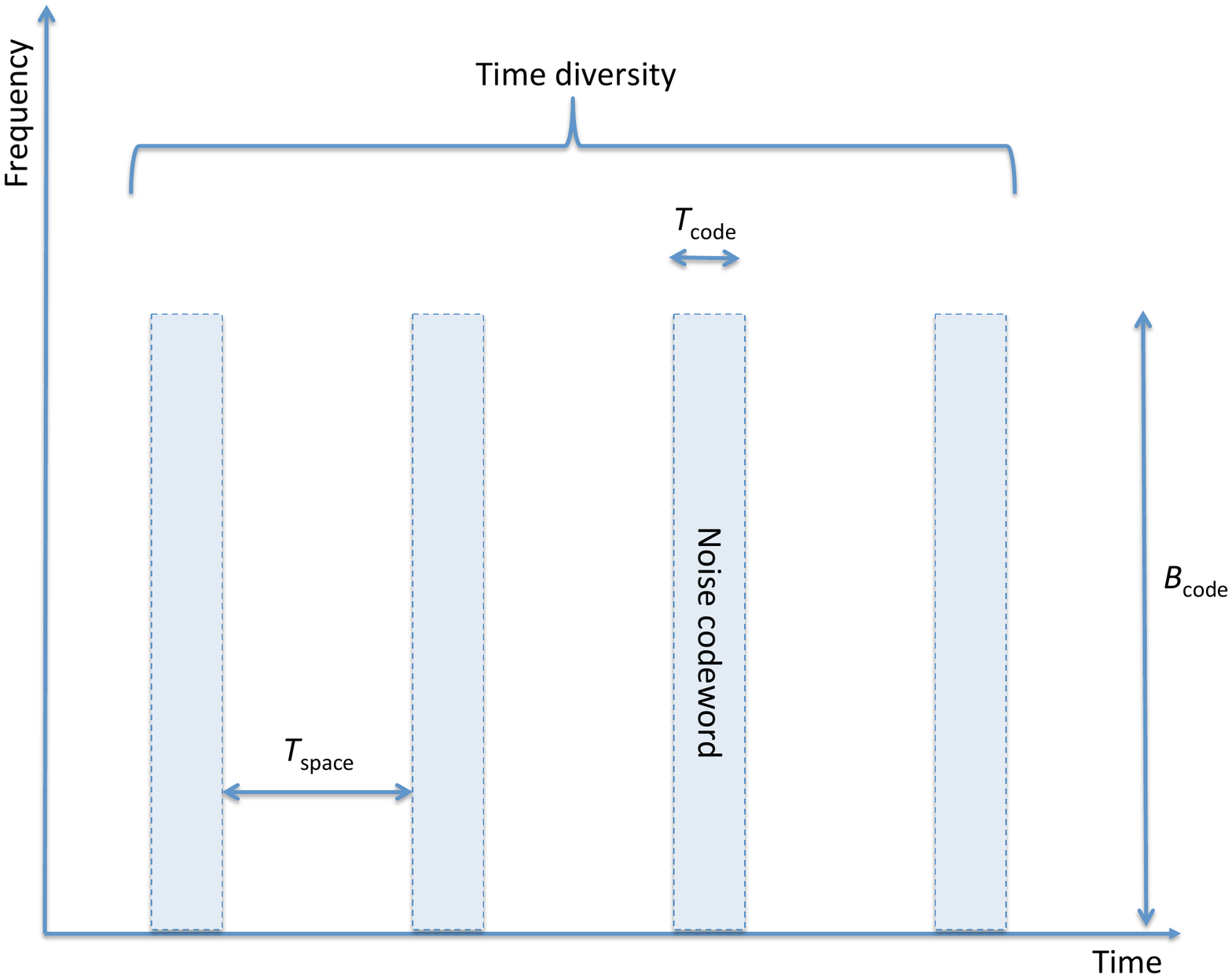}
	{1}
	{Illustration of time diversity. Each codeword is replicated $J$ times
	($J=4$ is shown) with a spacing $T_{\text{space}}$ sufficiently large to
	encounter independent scintillation.}

\subsubsection{Outage strategy}

Time diversity can also work at the level of messages rather than codewords.
If multiple replicas of a message are transmitted to another civilization
because it is not known when they may be listening, as a side benefit
the resulting redundancy provides a crude form of time diversity.
Energy efficiency is inevitably sacrificed with ''hard'' decoding of individual replicas,
but can be improved with ''soft'' decoding in which reliability information is preserved.
This suggests an \emph{outage} strategy for dealing with scintillation
as an alternative to time diversity \citeref{656}.

\section{Discovery}

Communication isn`t possible until the receiver discovers the signal.  
Discovery of a signal such Fig. \ref{fig:centauriDreams} 
\begin{changea}
can be based on the detection of individual energy bursts
followed by an appropriate pattern recognition,
even without knowledge of the overall signal structure
(such as  $\tcode$ and $\bcode$).
Statistical evidence for an information-bearing signal accumulates
with the detection of multiple bursts over a range of nearby frequencies. 
This signal is easy to distinguish from natural phenomena as well as artificial signals of terrestrial origin
(since sparsity in both time and frequency is atypical)
and bursts can be detected reliably
(since they must be energetic if information is to be extracted reliably).
\end{changea}

Discovery is hard in another respect, since due to time and frequency sparsity the receiver must be patient and conduct multiple observations in any particular range of frequencies to confidently rule out the presence of a signal with this character \citeref{656}.
A search over time and carrier frequency is required, as is a search over the
time scale of $h(t)$.
Estimates of $\tcoherence$ and $\bcoherence$ substantially reduce the scope of the search.
\begin{changea}
Many existing SETI search algorithms will associate this type of signal with false positive detections,
since it does not display the time and frequency persistence often used to distinguish 
a credible signal from a false positive.
\end{changea}

For communication with a civilization,
discovery needs to be taken into account in the design of the channel coding.
Channel coding that uses multiple levels
(more complicated than than the ''on'' and ''off'' energy bursts of the location code)
renders detection of bursts
less reliable and increases the required observation time.
Time diversity reduces the energy of individual bursts, making an even larger
impact on detection reliability.
Thus, an simple location code combined combined with an outage strategy
to deal with scintillation is particularly attractive
when the challenge of discovery is taken into account \citeref{656}.

\section{Conclusions}

End-to-end design of a digital communication system at interstellar distances has been considered,
with an emphasis on minimizing the energy per bit or energy per message delivered to the receiver. 
A solution with all the desired properties is identified.
\begin{changeb}
A location code is conceptually simple, and a signal of this type is straightforward
to discover at the receiver using a strategy based on detection of individual energy bursts
followed by pattern recognition.
\end{changeb}
This design can also approach the bandwidth-unconstrained limit on received energy,
even in the presence of scintillation (when it is combined with time diversity).
It does, however, require a significant bandwidth expansion.
A similar approach can also achieve energy-efficient information-free beacons \citeref{656}.

Perhaps most significantly, the fundamental principle of minimizing received energy
without bandwidth constraint and
in consideration of jointly observable ISM and motion impairments leads to
a simple and highly-constrained signal design.
One can hope that the transmitter and receiver designers addressing this joint challenge
without the benefit of coordination might arrive at compatible conclusions.
Interstellar impairments are fortuitous for uncoordinated communication since they constrain
the signal structure and parameterization.
The tighter constraints at lower carrier frequencies and at radio
as opposed to optical wavelengths are also helpful in this regard,
because they more tightly constrain the signal
without adversely impacting the energy requirements.

It is reasonable to ask two skeptical questions.
First, is it likely that another civilization is aware of the opportunity to reduce received energy,
and aware of the fundamental limit on received energy?
The history of earthbound communications leads to optimism.
Energy-limited communication near the fundamental limit was 
understood much earlier (in the 1950`s) than bandwidth-limited communication
(in the 1990`s) because of the simplicity of the solution.
Optical communication is typically much simpler than radio precisely because bandwidth has never been 
considered a limiting resource at the shorter optical wavelengths.

Second, is another civilization likely to be motivated by and act upon the opportunity to reduce the
received energy?
This is a more difficult question, since a more advanced civilization may well have
tapped into cheaper sources of energy.
Even if so, they may be motivated by the simplicity of the solutions and the benefits
of that simplicity to implicit coordination.
In addition,
even if energy is more plentiful, 
there are many beneficial ways to consume more energy other than deliberate inefficiency. 
They could increase message length, reduce the message transmission time, transmit in more directions simultaneously, or transmit to greater distances.
Overall it is unlikely that a civilization would use more energy than necessary unless for some reason they
consider a reduction in bandwidth to be a higher priority.

For communication with a starship, 
trading for greater bandwidth remains useful as a way to minimize received energy,
which is a particular benefit on the downlink from the starship because of the reduction in
transmit power and/or transmit antenna size.
Because there is the luxury of joint design of transmitter and receiver,
simplicity is not a goal in itself and if desired the bandwidth expansion can be reduced
without a significant penalty in received energy using something considerably
more complex than the location code.

\section*{Acknowledgements}

Early phases of this research were supported in part by a grant from the
National Aeronautics and Space Administration to the SETI Institute.
Ian S. Morrison of the Australian Centre for Astrobiology has maintained an invaluable ongoing
dialog on these issues.
\begin{changea}
James Benford of Microwave Sciences offered many valuable suggestions on
how to communicate with the intended audience of physicists and astronomers.
\end{changea}
The following individuals participated in many
discussions of issues surrounding interstellar communication:
Gerry Harp and Jill Tarter of the SETI Institute, and
Andrew Siemion and Dan Werthimer of the Space Sciences Laboratory at Berkeley.
Samantha Blair of the Joint ALMA Observatory was an early collaborator on
interstellar scattering, and was assisted by William Coles of the University of California at San Diego.
David Tse (now at Stanford University)
pointed out literature pertinent to power-efficient design.

\appendices 

\section{Model for interstellar propagation}
\label{sec:ismModel}

The impairments introduced in the ISM and due to motion are considered separately, one at a time.

\subsection{Signal and detection}
\label{sec:signal}

In the absence of any ISM impairments, assume that an energy burst is
represented by a passband waveform of the form
\begin{equation}
\label{eq:txsignal}
x(t) =2 \, \Re \left\{ \sqrt{\energy}  \, h(t) \, e^{\,i \, ( 2 \pi f_c \, t +  \theta)} \right\}
\end{equation}
where $\Re$ denotes the real part,
$\energy$ is the total energy of a burst at baseband, 
and $f_c$ Hz is frequency of a carrier with unknown phase $\theta$.
If $h(t)$ is confined to bandwidth $0 \le f \le B$ then $x(t)$ is confined to $f_c \le | f | \le f_c+B$.
The autocorrelation $R_h (t)$ of $h(t)$ is defined as $R_h (t) = h(t) \convolve \cc h(t)$ (where $R_h (0) = 1$)

\subsubsection{Demodulation}

Neglecting a double-frequency term, demodulation recovers a baseband signal
\begin{equation}
\label{eq:demod}
y(t) = e^{-\,i\,2 \pi f_c \, t} x(t) = e^{\,i\,\theta} \sqrt{\energy} \, h(t) \,.
\end{equation}
The matched filter of \eqref{eq:MF} 
eliminates the double-frequency term and
recovers signal $z(t) = e^{\,i\,\theta} \, \sqrt{\energy} \, R_h (t)$.
The signal component of a 
phase-incoherent energy estimate
as described in Sec. \ref{sec:phaseIncoherentDetection} is $|z(0)|^{\,2} = \energy$.

\subsubsection{Additive noise}
\label{sec:noiseModel}

Assume a real-valued AWGN $N(t)$ with power spectral density $\npsd$ is input to the receiver.
Since $h(t)$ has unit energy, the resulting $z(0) = N$ in \eqref{eq:channelModel}
is a complex-valued circular Gaussian random variable with
$\expec{N} = \expec{N^{\,2}} = 0$ and $\expec{|N|^{\,2}} = \npsd$.
No other ISM and motion impairments influence these noise statistics.

\subsubsection{Codeword detection}

Suppose \eqref{eq:demod} followed by a matched filter is repeated for every frequency where an
energy burst may reside in accordance with the basis shown in Fig. \ref{fig:codewordInternalStructure}.
Sampling those matched filter outputs at the appropriate times
yields an $M$-dimensional vector $\vec y$.
The optimum processing of $\vec y$ defined by \eqref{eq:receptionVector} chooses the signal
that is closest in Euclidean distance to $\vec y$, or 
specifically the $m$ for which $\big|| \vec y - \vec c_m \big||$ is minimum.
If $\big|| \vec c_m \big|| = \sqrt{\energy}$ does not depend on $m$,
this is equivalent to choosing the $m$ that maximizes $\Re \{ \hc{\vec c}_m \vec y \}$.
If each location has an unknown phase $\theta_m$ assumed to be
a uniformly distributed random variable, the optimum detector
chooses the maximium $\big| \, \hc{\vec c}_m \vec y \,  \big|$ instead \citerefWloc{43}{Sec.7.7}.
For the location codebook of \eqref{eq:cartesian} this criterion
reduces to choosing the coordinate of $\vec y$ with the largest magnitude.
The resulting bit error probability is bounded by \citeref{656}
\begin{equation}
\label{eq:peOrthogonal}
\pe \le \frac{1}{4} \, M^{\, 1 - \frac{\ebit/\npsd}{2 \, \log 2}} \,.
\end{equation}
If $\ebit / \npsd$ is greater than twice the limit of \eqref{eq:limit},
it follows that $\pe \to 0$ as $M \to \infty$.
There is a factor of two penalty due to phase-incoherent detection, but this can be circumvented by time diversity.

\subsection{Motion}
\label{sec:motionModel}

Neglecting relativistic effects, motion can be modeled as a changing 
distance between transmitter and receiver due to relative transmitter-receiver motion,
\begin{equation*}
d(t) = D + v \, t + \frac{a \, t^2}{2}
\end{equation*}
where $D$ is a fixed distance, $v$ and $a$ are the
 components of velocity and acceleration away from the source along the
line of sight.
The propagation delay then changes as $d(t) / c$, where $c$ is the speed of light.
This changing delay warps the
time axis of the passband signal \eqref{eq:txsignal}, resulting in
\begin{equation}
\label{eq:passbandDoppler}
2 \, \Re \left\{ \sqrt{\energy} \, h \left( t - \frac{d(t)}{c} \right)  \, 
e^{\,i \, ( 2 \pi f_c \, \left(t - \frac{d(t)}{c}  \right) +  \theta)}
 \right\}
\end{equation}
Of no consequence are the terms in $D$ and $v$.
In the case of a starship, $v$ is likely to be accurately known and is easily compensated,
and in the case of a civilization there is no prior coordination of either the time scale
or the carrier frequency so $v$ is merely another contributor to that unknown.

Assuming $D = v = 0$, following demodulation
of \eqref{eq:passbandDoppler}, the baseband signal is
\begin{equation*}
e^{- i \, 2 \pi \, a \, t^{\,2} / 2 \lambda_c } \cdot
h \left( 
		t- \frac{a \, t^{\,2} }{2 \, c ~} 
	\right) \,.
\end{equation*}
Acceleration thus causes two distinct impairments.

\subsubsection{Time-varying phase}

A quadratic phase shift of the carrier results in a
total phase variation significantly less than $\pi$ radians
if the time duration $T$ of $h(t)$ satisfies \eqref{eq:accelphase}.

\subsubsection{Time warping}

The quadratic time warping of $h(t)$ can't be studied by frequency decomposition.
Rather, the sampling theorem decomposes $h(t)$ into basis functions in time,
and those basis functions are differentially delayed with a total delay spread
\begin{equation*}
\Delta \tau = \frac{a \, T^2}{2 \, c ~} \le \frac{1}{2 \, f_c} \,.
\end{equation*}
This bound on $\Delta \tau$ assumes that \eqref{eq:accelphase} is already satisfied.
It will always be the case that $B \ll f_c$, 
and \eqref{eq:freqCoherence} is satisfied in that case.
Thus \eqref{eq:accelphase} always defines the acceleration coherence time.

\subsection{Interstellar dispersion and scattering}
\label{sec:groupDelay}

Although there will be variations in electron density
and phase shift in the scattering screen model of Fig. \ref{fig:scatteringIllus}
due to the turbulence of interstellar gas clouds,
these effects are slow relative to the time duration $T$ of an energy burst.
Neglecting the time variation due to this turbulence,
dispersion and scattering can be 
represented by a frequency response $G(f)$
at passband, with corresponding impulse response $g(t)$.
Referenced to baseband these are
$G(f+f_c)$ and $g(t) \, e^{\,-i\, 2 \pi \, f_c \, t}$.
After matched filtering, since convolution is commutative,
\begin{align}
\notag
z(t) =&
\Bigg ( h(t) \convolve \left( g(t) \, e^{\,-i\, 2 \pi \, f_c \, t} \right) \Bigg) \convolve \cc h(t) \\
\label{eq:dispersionModel}
=& R_h (t) \convolve \bigg( g(t) \, e^{\,-i\, 2 \pi \, f_c \, t} \bigg) \,.
\end{align}
The overall effect at the matched filter output is
 the dispersion applied to autocorrelation $R_h (t)$.
 Any dispersive smearing of the autocorrelation is not directly relevant,
but what does matter is any impairments that cause $\big| z(0) \big| \ll R_h (0) = 1$.

\subsubsection{Group delay}

For $G(f) = e^{\, i \, \phi (f)}$
the group delay $\tau (f)$ is defined as
\begin{equation}
\label{eq:groupDelay}
\tau (f) = - \frac{1}{2 \pi} \, \frac{\diff \, \phi (f)}{\diff  f} \,.
\end{equation}
For example, the Fourier transform of $h(t - \tau)$ is
$H(f) e^{\,- i \, 2 \pi \, f \, \tau}$, and thus $\phi (f) = - f \, \tau$ and $\tau (f) = \tau$,
or in this case the group delay is equal to the fixed delay $\tau$.

Assume the minimum and maximum group delays are
$\tau_{\text{min}}$ and $\tau_{\text{max}}$ respectively.
The total variation in group delay across bandwidth $B$ 
is the delay spread $\Delta \tau = \tau_{\text{max}} - \tau_{\text{min}}$.
For a given bandwidth $B$, a sufficiently
small $\Delta \tau$ has a negligible effect.
To see this, integrate \eqref{eq:groupDelay} to recover
the phase from the group delay,
\begin{equation*}
\phi (f) =  \phi(0) - 2 \pi \, f \, \tau_{\text{min}} - 2 \pi \, \int_{0}^{f} \left( \tau (u) - \tau_{\text{min}} \right)  \, \diff u
\,.
\end{equation*}
The receiver will be explicitly designed to be impervious to the first two terms,
and the remainder is bounded by
\begin{equation*}
\big| \, \phi (f) - \phi (0) + 2 \pi \, f \, \tau_{\text{min}} \, \big| \le  2 \pi \, f \, \Delta \tau \le 2 \pi \, B \, \Delta \tau \,.
\end{equation*}
Thus the deleterious portion
of $\phi (f)$ is guaranteed to be small relative to $2 \pi$
when \eqref{eq:freqCoherence} is satisfied.

\subsubsection{Plasma dispersion}
\label{sec:plasmaDispersionModel}

If $G(f) = e^{\,i\,\phi(f)}$ then the Fourier transform of \eqref{eq:dispersionModel}
is $| H(f) |^2 \, e^{\,i\,\phi(f+f_c)}$.
The Taylor series expansion of $z(0)$ for small $\phi (f)$ is
\begin{align*}
\int_{0}^{B} |H(f)|^2 \, &  e^{\,i\,\phi(f+f_c)} \, \diff f \approx \\
&\int_{0}^{B} |H(f)|^2 \, \left( 1 - \frac{\phi^2 (f+f_c)}{2} \right) \, \diff f \\
&+ i  \int_{0}^{B} |H(f)|^2 \,  \phi (f+f_c) \,\, \diff f \,.
\end{align*}
There is some reduction in the real part of the signal for $\phi(f) \ne 0$,
as well as some leakage of the autocorrelation into the imaginary part.
Any detrimental reduction in the energy estimate $\big| z(0) \big|$ is avoided if
$| \phi (f + f_c) |$ is small.

\subsubsection{Scattering}
\label{sec:geometry}

When dealing with different scales on the diffraction screen,
a convenient reference is the Fresnel scale $\displaystyle r_F = \sqrt{\lambda_c \, D}$,
typically on the order of light-seconds.
The antenna aperture diameter $R \ll r_F$, and for strong scattering
$r_F$ is much larger than the diffraction scale.

For the geometry of Fig. \ref{fig:diffractionZone},
consider the condition on $\Delta$ such that the
difference in propagation distance from
the receiver to point $x$ and to
point $x+\Delta$ is equal to a half wavelength $\lambda /2$.
This is the largest $\Delta$ for which there can be 
strong constructive interference at the receiver
for a plane wave impinging on this patch.
By the far field approximation (valid when $D \gg |x|$),
the two triangles are congruent, $\theta \sim 0$ and
$\tan \theta \sim \sin \theta \sim \theta$,
and it follows that $\Delta \sim \lambda_c D / (2 \, x)$
and further  the excess delay from point $x$ is $\tau \sim x^2 / ( D \, c )$.
Eliminating $x$,
\begin{equation*}
\frac{\Delta}{r_F} \sim \frac{1}{\displaystyle 2 \sqrt{f_c \tau}} \,.
\end{equation*}
Since $\Delta \to 0$ as $\tau \to \infty$
(regardless of the statistics of the turbulent plasma)
the net receive energy is dominated by the geometry,
and the plane-wave
energy impinging on the scattering screen that has an opportunity to
constructively interfere at large group delay $\tau$ approaches zero.

\incfig
	{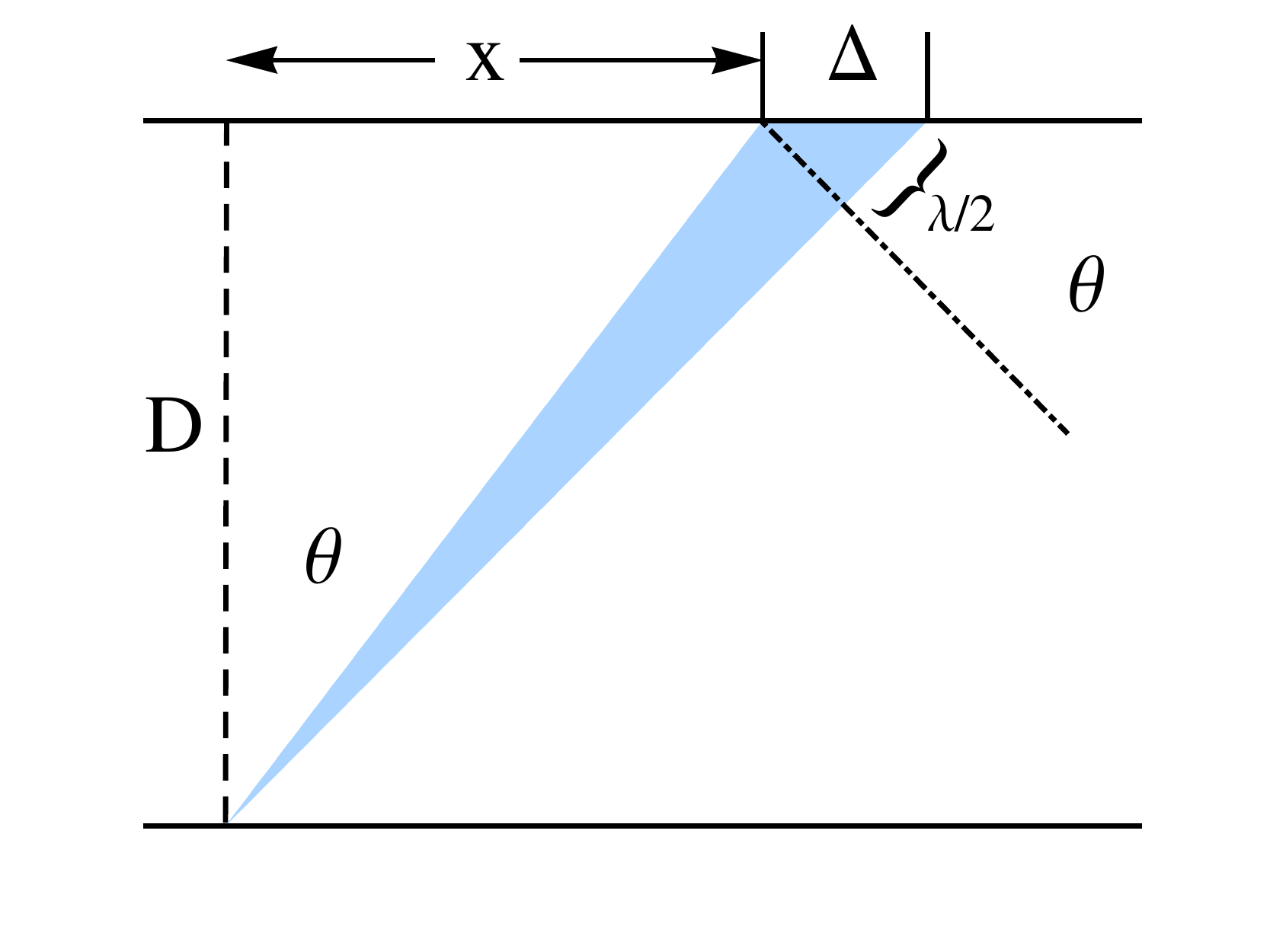}
	{.6}
	{The geometry representing the maximum range of constructive interference
	from a single coherent patch on the scattering screen.}

The main lobe of the transmit and receive antennas will be the ultimate limitation on $\tau$.
For an antenna with aperture $R$,
using the same far-field approximation the main lobe referenced to the scattering screen
is $|x| < x_R$ where
\begin{equation*}
\frac{x_R}{r_F} \sim \frac{r_F}{2 \, R} \,.
\end{equation*}
It will always be the case that $r_F \gg R$, and thus 
that portion of the scattering screen that contributes to scattering falls well within
the main lobe of both the transmit and receive antennas.

\end{makefigurelist}  

\bibliographystyle{IEEEtran}
\bibliography{SETI-references,patents}

\begin{IEEEbiographynophoto}{David G. Messerschmitt}
is the Roger A. Strauch Professor Emeritus of Electrical Engineering and Computer Sciences (EECS) at the University of California at Berkeley. At Berkeley he has previously served as the Chair of EECS and the Interim Dean of the School of Information. He is the co-author of five books, including \emph{Digital Communication} (Kluwer Academic Publishers, Third Edition, 2004). He served on the NSF Blue Ribbon Panel on Cyberinfrastructure and co-chaired a National Research Council (NRC) study on the future of information technology research. His doctorate in Computer, Information, and Control Engineering is from the University of Michigan, and he is a Fellow of the IEEE, a Member of the National Academy of Engineering, and a recipient of the IEEE Alexander Graham Bell Medal recognizing ``exceptional contributions to the advancement of communication sciences and engineering''.
\end{IEEEbiographynophoto}

\end{document}